\documentclass[journal]{vgtc}                     

\usepackage{graphicx}

\newcommand{\revise}[1]{\textcolor{black}{#1}}

\vgtccategory{Research}

\vgtcpapertype{algorithm/technique}

\title{\tool{}: Graph Simplification via Adaptive Motif Design}

\author{%
Hong Zhou, Peifeng Lai, Zhida Sun, Xiangyuan Chen, Yang Chen, Huisi Wu, Yong Wang
}

\authorfooter{
 \item
 	H. Zhou, P. Lai, Z. Sun, X. Chen, Y. Chen and H. Wu are with Shenzhen University, Shenzhen, China. E-mail: hzhou@szu.edu.cn, laipeifeng1111@gmail.com, zhida.sun@szu.edu.cn, \{2310274034, 2310273092\}@email.szu.edu.cn, hswu@szu.edu.cn. 
  \item
  	Y. Wang is with Nanyang Technological University, Singapore. Part of this work was done when he was affiliated with Singapore Management University. E-mail: yong-wang@ntu.edu.sg. Y. Wang is the corresponding author.
}

\abstract{%
With the increase of graph size, it becomes difficult or even impossible to visualize graph structures clearly within the limited screen space. Consequently, it is crucial to design effective visual representations for large graphs.
In this paper, we propose \tool{}, a novel approach that can capture the essential structure patterns of large graphs and effectively reveal the overall structures via adaptive motif designs.
Specifically, our approach involves partitioning a given large graph into multiple subgraphs, then clustering similar subgraphs and extracting similar structural information within each cluster. 
Subsequently, adaptive motifs representing each cluster are generated and utilized to replace the corresponding subgraphs, leading to a simplified visualization.
Our approach aims to preserve as much information as possible from the subgraphs while simplifying the graph efficiently.
Notably, our approach successfully visualizes crucial community information within a large graph.
We conduct case studies and a user study using 
real-world graphs to validate the effectiveness of our proposed approach. The results demonstrate the capability of our approach in simplifying graphs while retaining important structural and community information.
}


\teaser{
  \centering
\begin{tabular}{c}
 \includegraphics[width=6.25in]{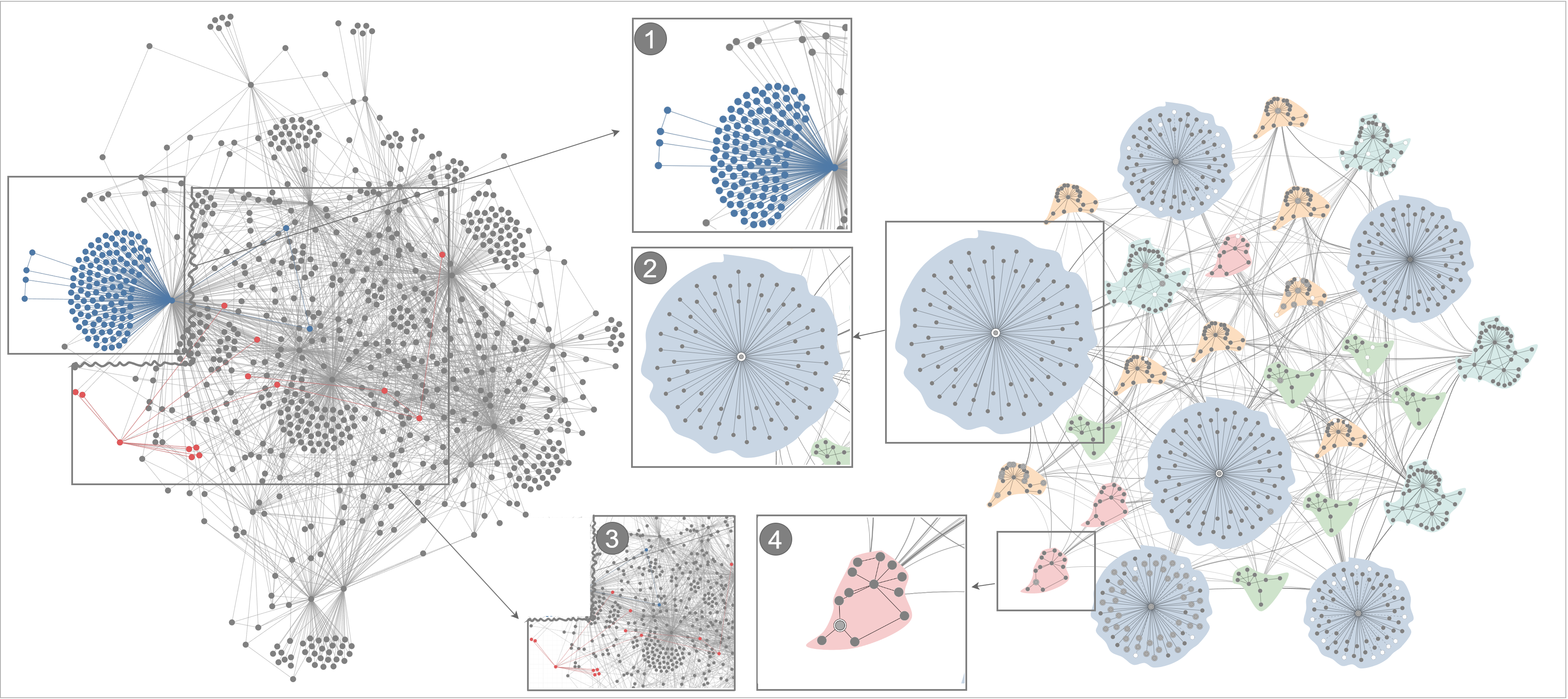}\\
\end{tabular}
\begin{tabular}{cc}
 \multicolumn{2}{c}{(a) \hspace{3.3in} (b)} 
\end{tabular}
  \caption{
  Case analysis of the Cpan dataset~\cite{heymann2009cpan}: (a) the original graph; (b) our simplified graph. The highlighted areas of each subfigure show the enlarged communities \textcolor{black}{labeled as ``1'' - ``4''. The communities labeled as ``1'' and ``2'' are the same, and ``3'' and ``4'' are the same. To make communities ``1'' and ``3'' in (a) easier to identify, their nodes and edges are highlighted in blue and red, respectively. To facilitate the comparison, communities ``1'' and ``2'' are aligned vertically, and ``3'' and ``4'' are aligned horizontally. In (b), motifs with the same color and similar shape represent similar communities. The size of the motif indicates the number of nodes in this community. For more details on our visual encoding scheme, please refer to Sec. 3.} Our result provides a clearer expression of community information. 
  }
  \label{fig:Cpan-teaser}
}

\graphicspath{{figs/}{figures/}{pictures/}{images/}{./}} 

\usepackage{tabu}                      
\usepackage{booktabs}                  
\usepackage{lipsum}                    
\usepackage{mwe}                       

\usepackage{mathptmx}                  
\usepackage{multirow}

\usepackage{url}
\usepackage{amssymb}

\newcommand{\tool}{\textit{AdaMotif}}

\begin{document}



\firstsection{Introduction}

\maketitle


The graph, a common form of relational data, consists of nodes and edges. 
Examples of graphs include social networks, network topologies, and molecular structures. As the scale of graphs increases, analyzing large-scale graphs pose significant challenges in both data visualization and data mining domains. 

In graph visualization, the node-link diagram method is commonly used due to its intuitive and effective representation. 
However, as the scale of graphs increases, corresponding node-link diagrams become more large-scale, and the structural information within the graphs also becomes more complex. Effective visualization methods can aid users in analyzing the structural information within large graphs. One feasible approach is to simplify large-scale node-link diagrams.

Currently, research on simplifying large graphs mainly focuses on two directions: glyph-based simplification (e.g. motif simplification~\cite{dunne2013motif}) and graph sampling~\cite{jiao2022hierarchical, zhao2020preserving, zhou2020context}). Glyph-based simplification abstracts the graph structure, whereas graph sampling reduces the number of nodes to achieve simplification.
Motif simplification can highlight special graph structures by representing them with motifs that require less screen space. Therefore, it can be used for analyzing large graphs and has a wide range of practical applications~\cite{dunne2013motif}, e.g., social media networks~\cite{NodeXL}. However, these motifs are predefined and can only be applied to three specific structures (i.e., fan, connector, and clique) rather than general structures. Graph sampling is used to simplify the analysis and processing of large-scale networks by working with a manageable subset of the data. This technique can reduce the graph size but cannot emphasize the community structures in the graph.

Community information in graphs has significant practical applications. For example, in social networks, community information can be used to effectively monitor the development of public opinion and recommend suitable products to users~\cite{cabrera2021influence}. 
By comparing the similarities and differences between communities, more precise policies can be formulated and implemented, enhancing policy effectiveness. Therefore, extracting and analyzing community structures, as well as the similarities and differences between them, is of great importance in areas such as social network analysis~\cite{majeed2020social,tabassum2018social}, information retrieval~\cite{hoeber2018information}, recommendation systems~\cite{guo2020survey, zhou2020interactive}, transportation and logistics~\cite{gupta2020developing}, and biological network analysis~\cite{doncheva2012topological}.

In the field of graph mining, general community structures in graphs can be detected by subgraph partitioning~\cite{de2011generalized, zeng2018distributed} and graph embedding~\cite{narayanan2017graph2vec, rozemberczki2020characteristic}, and can be compared by graph alignment~\cite{chen2020cone, heimann2021refining, hermanns2023grasp, nassar2018low}. These methods effectively assist users in mining community information from graphs but lack intuitive expression, which hinders users from further exploration and understanding. 

We integrate graph visualization with related graph mining techniques to propose a novel method for large graph simplification via adaptive motif design. Our approach first divides a graph into subgraphs (i.e., communities) based on their structural information, resulting in numerous subgraph pieces. Then, it clusters similar subgraphs together and calculates the centroid subgraph for each cluster, which represents the structural information of that class of subgraphs. Finally, it aligns the representative subgraphs to obtain their similar structures, thereby generating a simplified motif that represents the overall structure of each subgraph. By replacing the original subgraphs with simplified motifs, the visual complexity of the visualization for the original graph is significantly reduced, greatly improving visual clarity and readability. 
Similar to other graph simplification methods, our approach 
also results in
information loss, which is inevitable. Although it loses the original edge information between nodes in different communities, our method is well-suited for analyzing the overall topology of communities and the similarities and differences between them. Furthermore, our method is not limited to analyzing specific community structures.
It can automatically generate adaptive motifs for general community structures to help users visually analyze the characteristics of graphs.

The major contributions of this paper are as follows:
\begin{itemize}
    \item A novel graph visualization simplification framework, \tool{}, where the simplification process generates adaptive motifs automatically, representing the graph structure.
    \item A novel approach for community similarity representation, ensuring a uniform layout effect among similar communities to facilitate the discovery of their similarities.
    \item A novel method for community difference representation, showcasing differences among communities with similar structures.
\end{itemize}


\section{Related Work}

\subsection{Graph Visualization Simplification}

Existing research on graph visualization simplification can be categorized into two groups: glyph-based simplification and graph sampling. Glyph-based simplification achieves simplification by abstracting the graph structure, while graph sampling achieves simplification by directly reducing the number of nodes.

\textbf{Glyph-based Simplification.} Using specialized glyphs to represent corresponding data is an effective visualization method that can be applied across different datasets~\cite{brehmer2021generative, fuchs2016systematic, ying2022metaglyph}. A motif is defined as a subgraph pattern that occurs more frequently than expected by chance in a given network~\cite{alon2007network}. Motif simplification~\cite{dunne2013motif} utilizes graphical representation to simplify the visualization of graphs by designing special motifs to enhance the readability of graphs, particularly those with three specific structures (i.e., fan, clique, and connector) rather than general structures. Apart from motif simplification, other simplification methods involve merging nodes based on node or edge weights~\cite{purohit2014fast, zhou2022topological, oliver2023scalable}. While effectively reducing the scale of graphs and enhancing crucial information, these approaches 
lead to the loss of significant structural information presented in the original graph.

\textbf{Graph Sampling.} Graph sampling methods are commonly used to reduce the scale of graphs,
and thus simplify the graph visualization~\cite{cui2022survey}. 
\revise{Designing effective sampling methods can retain the core important structures of the graphs while reducing visual clutter. The MCGS~\cite{zhao2020preserving} method samples a small number of special nodes to preserve the key structure of the graph. 
Other graph sampling methods include Context-aware~\cite{zhou2020context} and Hierarchicalg~\cite{jiao2022hierarchical} sampling, etc.} However, as the sampling rate decreases, more information from the graph is lost. 

While our method also experiences information loss, emphasizing community structures, similarities, and differences helps users better analyze the relationships between communities and the overall topology, leading to a deeper understanding of community dynamics.

\subsection{Graph Structure Analysis}

Graph mining involves using methods such as graph theory and network analysis to explore, mine, and understand patterns within graphs. 

\textbf{Graph Clustering.} Analyzing the community structure in graphs is one of the important tasks in graph mining. Common methods for this include subgraph partitioning, such as Louvain~\cite{de2011generalized} and Infomap~\cite{zeng2018distributed}.
Analyzing the similarity between two graphs can be achieved by computing the graph edit distance~\cite{gao2010survey}. 
The distance effectively helps us analyze the structural similarity between subgraphs and can also be used for subgraph clustering. Furthermore, graph embedding is another important method that transforms the structural information of a graph into vector representations~\cite{narayanan2017graph2vec, rozemberczki2020characteristic,grover2016node2vec}. Through graph embedding, we can map graphs into vector space and then use clustering algorithms (such as k-means~\cite{hartigan1979algorithm}, spectral clustering~\cite{von2007tutorial}, etc.) to cluster graph vectors, thereby clustering similar graphs. 

\textbf{Graph Alignment.} When users need to analyze the correspondence between nodes in two graphs, graph alignment~\cite{trung2020comparative, nassar2018low} is an effective method. Through graph alignment, we can calculate the graph similarity matrix between two graphs, obtaining the similarity of any nodes between the two graphs. Based on this similarity matrix, we can infer the correspondence between similar nodes, further understanding and analyzing the structure and relationships between the two graphs.

\textbf{Graph Summarization.} 
\revise{Descriptive analysis of graphs is also important~\cite{liu2018graph}. For instance, the VoG method~\cite{koutra2015summarizing}
is a large-scale graph analysis approach based on the maximum description principle. This method extracts the most descriptive subgraph structures from a large graph to help users understand its overall characteristics.} On the other hand, the Momo method~\cite{coupette2021graph} focuses on analyzing and describing the similarities and differences between two or more graphs. This method can compare features such as structures, node attributes, and edge relationships between different graphs, helping users discover commonalities and differences between graphs. Furthermore, the SSumM method~\cite{lee2020ssumm} generates sparse summary graphs, summarizing the information of the graph to save storage space and analysis time.

Graph structure analysis methods can help users uncover important information, but their results are often not intuitive. Our approach employs relevant graph mining methods to extract communities and identify their similarities and differences, and designs effective visualization techniques to help users intuitively view and understand them.

\begin{figure*}[t]
  \centering
  \includegraphics[width=0.92\textwidth, keepaspectratio=true]{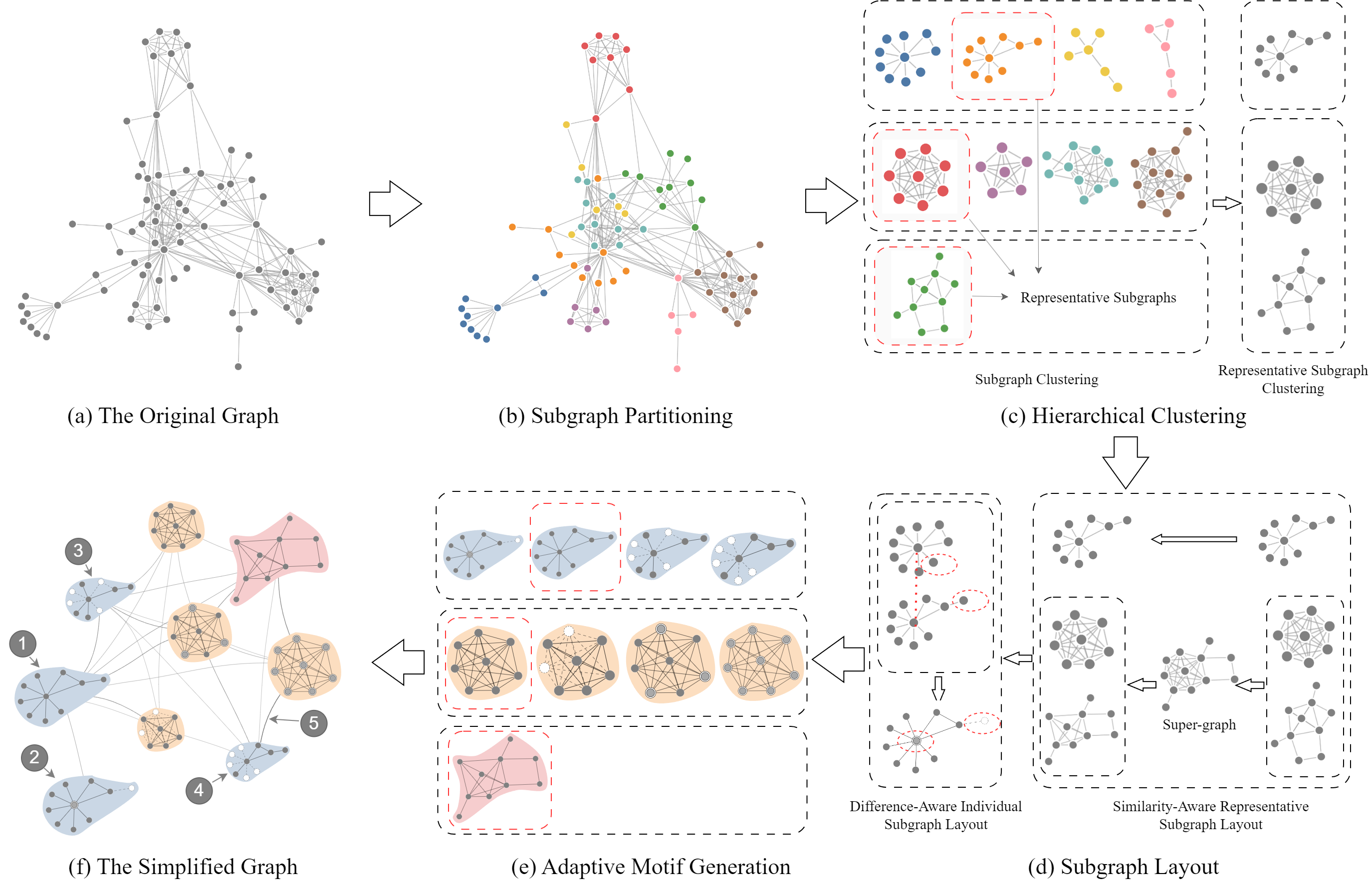} 
  \caption{Our \tool{} framework overview: (a) the original force-directed layout graph~\cite{d3-component}; (b) partitioning (a) into subgraphs annotated with different colors; (c) clustering subgraphs into categories indicated by gray dashed boxes, selecting a cluster center as the representative subgraph for each category, and further clustering these representatives into categories indicated by gray dashed boxes; (d) laying out representative subgraphs based on the super-graph to show similarities and then laying out individual subgraphs to show differences. The red dashed ovals highlight the graph differences and their encoding in our results; (e) generating adaptive motifs with colors encoding the categories; (f) the final simplified graph.}
  \label{fig:workflow}
\end{figure*}





\section{Adaptive Motif Design}
Our \tool{} is a novel graph simplification approach aimed at extracting and encoding graph structures into adaptive motifs, thus simplifying the graph and assisting users in quickly capturing graph community structures and understanding both the overall and local graph structures.


Fig.~\ref{fig:workflow} illustrates the framework of our approach. 
The original force-directed layout graph, with 77 nodes and 254 edges, is shown in Fig.~\ref{fig:workflow}(a). This graph is divided into nine subgraphs, each annotated with a different color in Fig.~\ref{fig:workflow}(b). These nine subgraphs are clustered into three categories indicated by gray dashed boxes shown in the left part of Fig.~\ref{fig:workflow}(c), and each category is selected with a representative subgraph, highlighted with a red dashed box. These three representative subgraphs are clustered into two categories, indicated by gray dashed boxes shown in the right part of Fig.~\ref{fig:workflow}(c). For the representative subgraph category with two graphs, a super-graph is synthesized for further visually representing their similarities in the layout as shown in the right part of Fig.~\ref{fig:workflow}(d). After fixing the layouts for all representative subgraphs, each non-representative subgraph in Fig.~\ref{fig:workflow}(c) is compared with its representative subgraph in order to generate its layout. Similar nodes are retained, while different nodes are either marked in white or gray with an outer ring. \revise{Nine motifs for each subgraph in Fig.~\ref{fig:workflow}(c) is generated and illustrated in Fig.~\ref{fig:workflow}(e). Finally, the sizes of the motifs are scaled according to the number of nodes in the respective subgraphs, and the final result (see Fig.~\ref{fig:workflow}(f)) is generated by performing an overall layout.} In the following subsections, we describe each step in detail.

\subsection{Subgraph Partitioning}

\revise{Community information is a key graph structure. 
We use a mature community detection algorithm~\cite{blondel2008fast} that automatically identifies community subgraphs without needing prior settings for the number or size of communities. Fig.~\ref{fig:workflow}(b) shows the partitioning result from Fig.~\ref{fig:workflow}(a), with nine subgraphs in different colors. In Fig.~\ref{fig:workflow}(b), while most communities are easily identifiable, the yellow, orange, and cyan ones are intertwined, making visual analysis challenging and highlighting the need for graph mining algorithms.}


\subsection{Hierarchical Clustering}

Our hierarchical clustering consists of two steps: first, clustering the
subgraphs, and then, for each subgraph category, identifying representative subgraphs and further clustering these representative subgraphs.

\textbf{Subgraph Clustering.} After detecting community subgraphs, we want to further present these subgraphs clearly and allow for intuitive observation of the similarities between them. For the subgraph similarity, we measure it using representative vectors computed by FEATHER~\cite{rozemberczki2020characteristic} and employ the Affinity Propagation method~\cite{frey2007clustering} to cluster subgraphs. Affinity Propagation aligns well with our objective of generating designs automatically, as it does not require specifying the number of clusters beforehand. As shown in Fig.~\ref{fig:workflow}(c), the nine subgraphs are clustered into three categories, each indicated by a dashed box. We can observe that there is a certain degree of structural similarity among the subgraphs within each category.

\textbf{Representative Subgraph Clustering.} For each subgraph category, we compute a representative subgraph based on the cluster center of this category. These representative subgraphs exhibit fewer differences from other subgraphs within the same category and are used for subsequent adaptive motif generation. Furthermore, we conduct another round of clustering on these representative subgraphs. The three representative subgraphs in Fig.~\ref{fig:workflow}(c) are further divided into two categories and indicated by gray dashed boxes. In the next subsection, we will rely on this clustering result to compute the layout of representative subgraphs, showcasing the local structural similarities between them.

\subsection{Subgraph Layout}

Our adaptive motif design is highly based on the representative subgraphs for each cluster. The adaptive motifs are used to further represent the original graph, and thus, the structure of the graph can be effectively simplified. However, this simplification may result in information loss. We found that there are certain similarities and differences between subgraphs in the same cluster and representative subgraphs in the same cluster. Therefore, we aim to express these similarities and differences in our design. 
Our subgraph layout has two components: a similarity-aware representative subgraph layout is designed to illustrate the similarities between representative subgraphs, and a difference-aware individual subgraph layout is designed to illustrate the differences between subgraphs in the same cluster.

\begin{figure}[t]
\centering
 \includegraphics[width=0.45\textwidth]{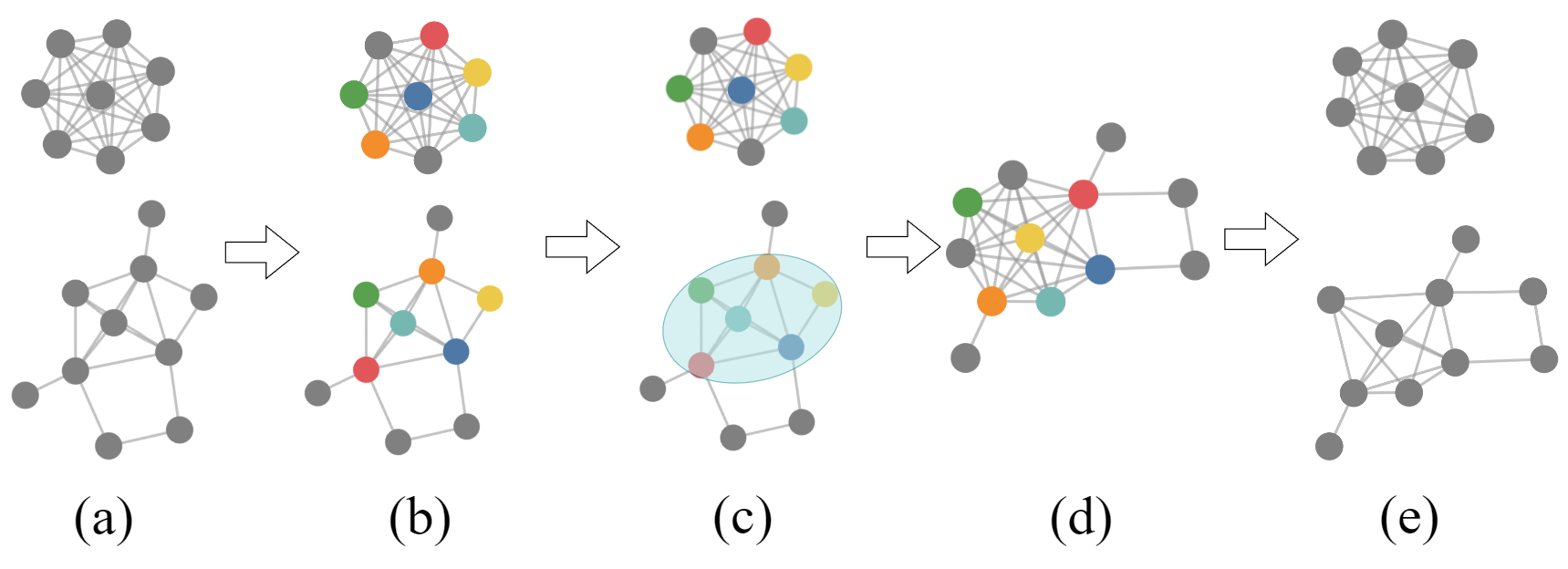}
  \caption{An example of the similarity-aware representative subgraph layout algorithm. Nodes with the same color indicate that they are matched through graph alignment. In (c), the yellow nodes on the periphery are placed in the center of the generated super-graph in (d). This is not due to their importance, but because the positions of the blue and yellow nodes were swapped. The blue nodes needed to be on the periphery to connect to the unmatched gray nodes. This did not significantly impact the final similarity-aware layout result in (e).
}
  \label{fig:supergraph}
\end{figure}

\textbf{Similarity-Aware Representative Subgraph Layout.} In the hierarchical clustering step, we classified the representative subgraphs into clusters, each with certain similarities. We need to extract the similarities and encode them into our layout. In the field of graph matching, the maximum common subgraph method can be used to accurately extract this graph similarity. However, this method is an NP-hard problem~\cite{hartmanis1982computers, kann1992approximability} and is only applicable to cases where the graphs are relatively small. Therefore, we opt for graph alignment methods~\cite{trung2020comparative, nassar2018low} to obtain inaccurate graph-matching results. Despite the presence of some errors, this method has mature algorithms and low time complexity, making it effective in detecting approximate graph similarity information. 

In our method, we employ the LREA~\cite{nassar2018low} method as a functional module for graph alignment. 
It outputs similarity values for each pair of nodes from two graphs. We prioritize matching node pairs with both high node degrees and high similarity values, while node pairs with low similarity values are ignored. Since graph alignment algorithms are only suitable for aligning two graphs at a time, for clusters with multiple representative subgraphs, we rely on a synthesized super-graph to achieve the alignment effect. 

For each representative subgraph cluster, if there is only one graph, we do not synthesize a super-graph (e.g., the cluster at the top of Fig.~\ref{fig:workflow}(d)). If there are multiple graphs (e.g., the cluster at the bottom of Fig.~\ref{fig:workflow}(d)), we select one graph as the basis and align another graph to generate the super-graph. Fig.~\ref{fig:supergraph} illustrates an example of the progress in generating the super-graph. In Fig.~\ref{fig:supergraph}(a), the two representative subgraphs belong to the same category, indicating a certain level of similarity between them. Therefore, we further use the LREA method to compute their aligned nodes (i.e., similar nodes) and color them in Fig.~\ref{fig:supergraph}(b). Nodes in the upper and lower graphs with the same color are aligned nodes, while the gray nodes are unaligned nodes. Subsequently, in Fig.~\ref{fig:supergraph}(c), we use the upper graph as a basis. The aligned nodes in the lower graph are removed, and the remaining unaligned nodes are then connected to the corresponding aligned nodes in the upper graph based on their connections to the removed aligned nodes, resulting in the super-graph shown in Fig.~\ref{fig:supergraph}(d). If there are other graphs within this representative subgraph category, the super-graph will be continuously updated as it aligns with other graphs one by one. Next, we apply force-directed node-link diagram layout~\cite{d3-component} to the super-graph to obtain layout information for all the nodes (see Fig.~\ref{fig:supergraph}(d)). Then, we align the node layouts of all the representative subgraphs with the corresponding node layouts of their super-graphs to illustrate their similar structures (see Fig.~\ref{fig:supergraph}(e)). As shown in Fig.~\ref{fig:supergraph}, the layout transformation of the two graphs from Fig.~\ref{fig:supergraph}(a) to Fig.~\ref{fig:supergraph}(e) facilitates the observation of their similarities. Therefore, this example of similarity-aware representative subgraph layout demonstrates an effective way to show graph similarity.

\begin{figure}[t]
\centering
 \includegraphics[width=0.5\textwidth]{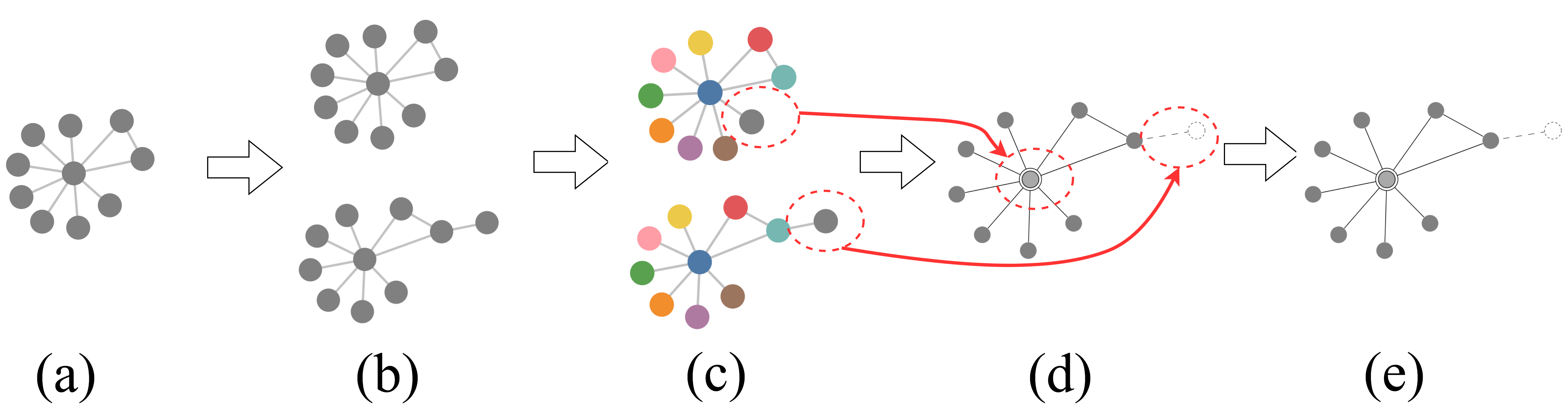}
  \caption{An example of the difference-aware individual subgraph layout algorithm. Nodes with the same color indicate that they are matched through graph alignment. The red dashed ovals highlight the graph differences in (c) and their encoding in (d).
}
  \label{fig:diff}
\end{figure}

\begin{figure}[t]
\centering
\centering
\begin{tabular}{ccccc}
 \includegraphics[width=0.118in]{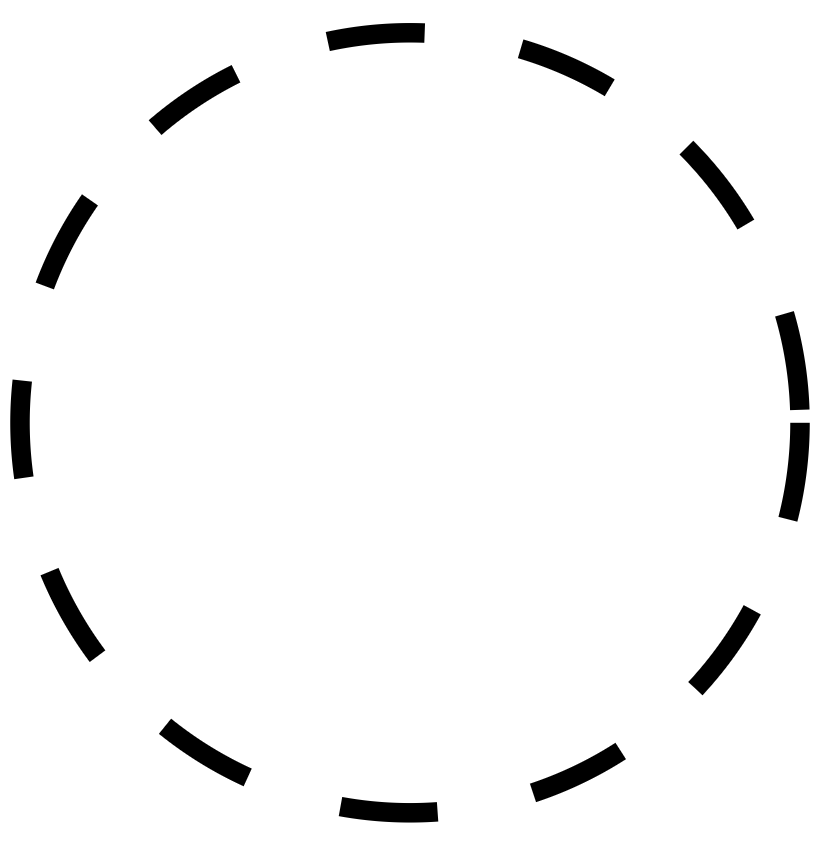} &  \includegraphics[width=0.12in]{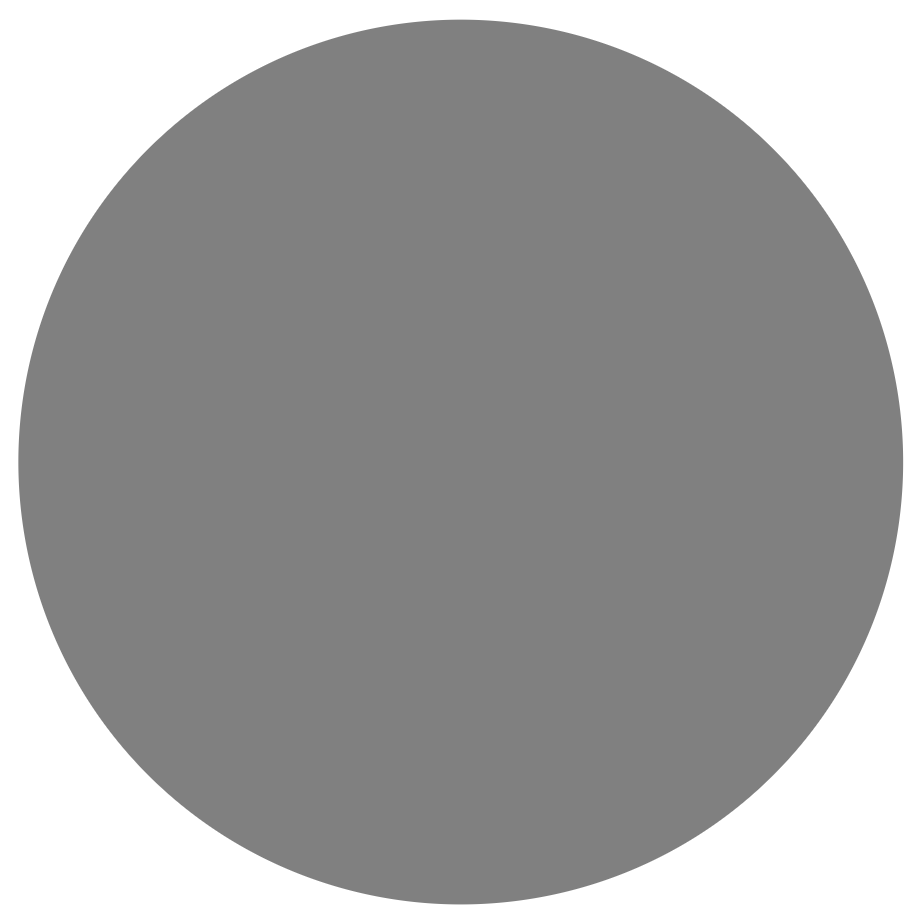} & \includegraphics[width=0.135in]{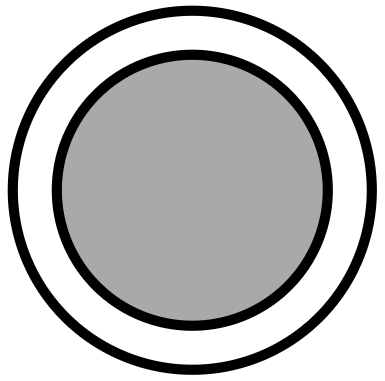} &  \includegraphics[width=0.18in]{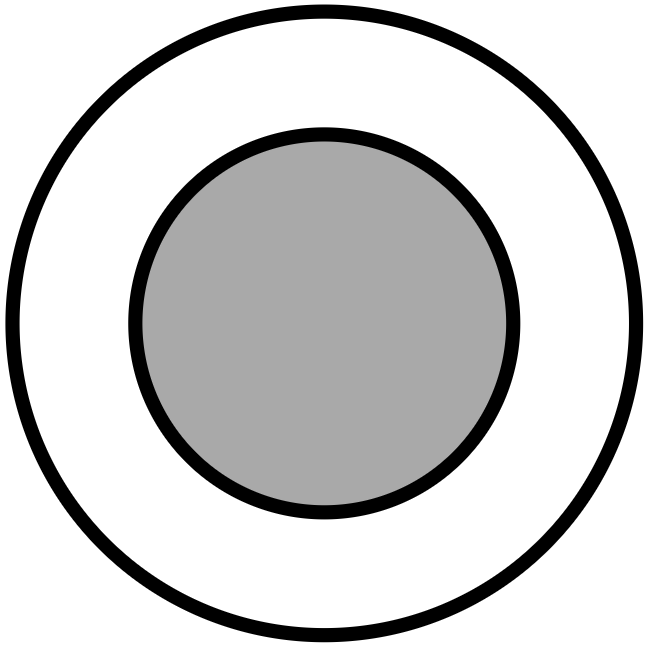} &  \includegraphics[width=0.23in]{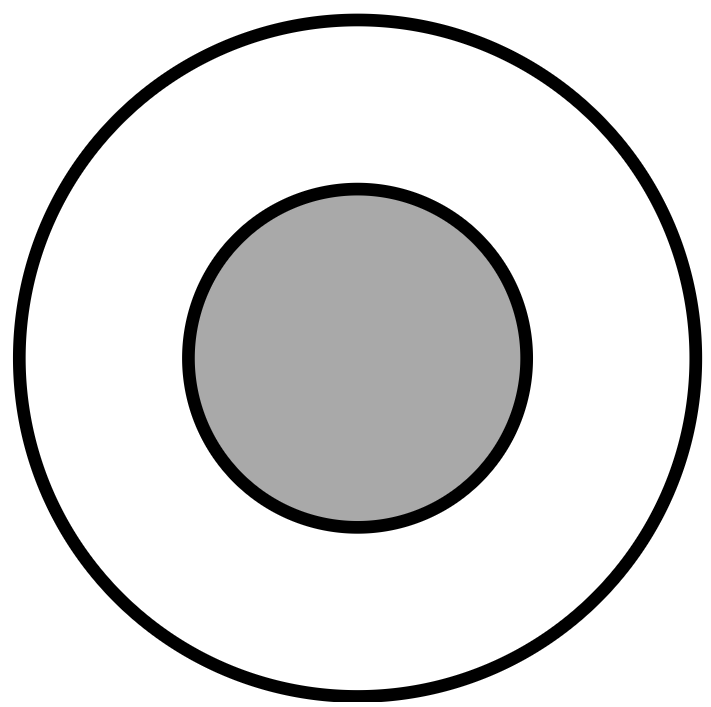}\\
(a) & (b) & (c) & (d) & (e)
\end{tabular}
  \caption{Five node encoding types in the difference-aware individual subgraph layout algorithm: (a) unaligned nodes; (b) aligned nodes not linked to unaligned nodes; (c)-(e) aligned nodes linked to unaligned nodes, and a wider outer ring indicates more linked unaligned nodes.
}
  \label{fig:encoding}
\end{figure}


\begin{figure*}[ht]
\centering
\begin{tabular}{cc}
\includegraphics[width=2.7in]{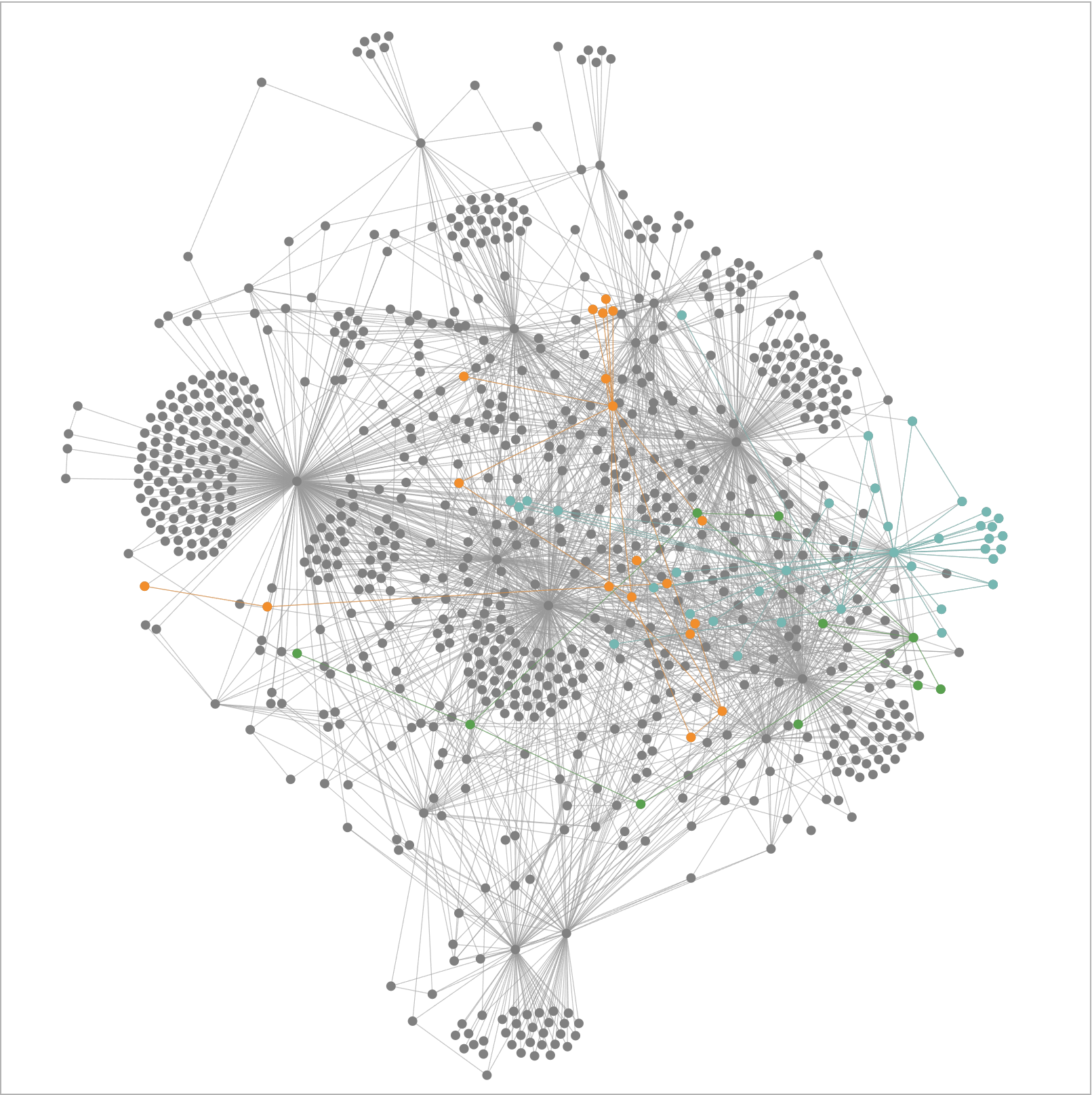} &  
\includegraphics[width=2.7in]{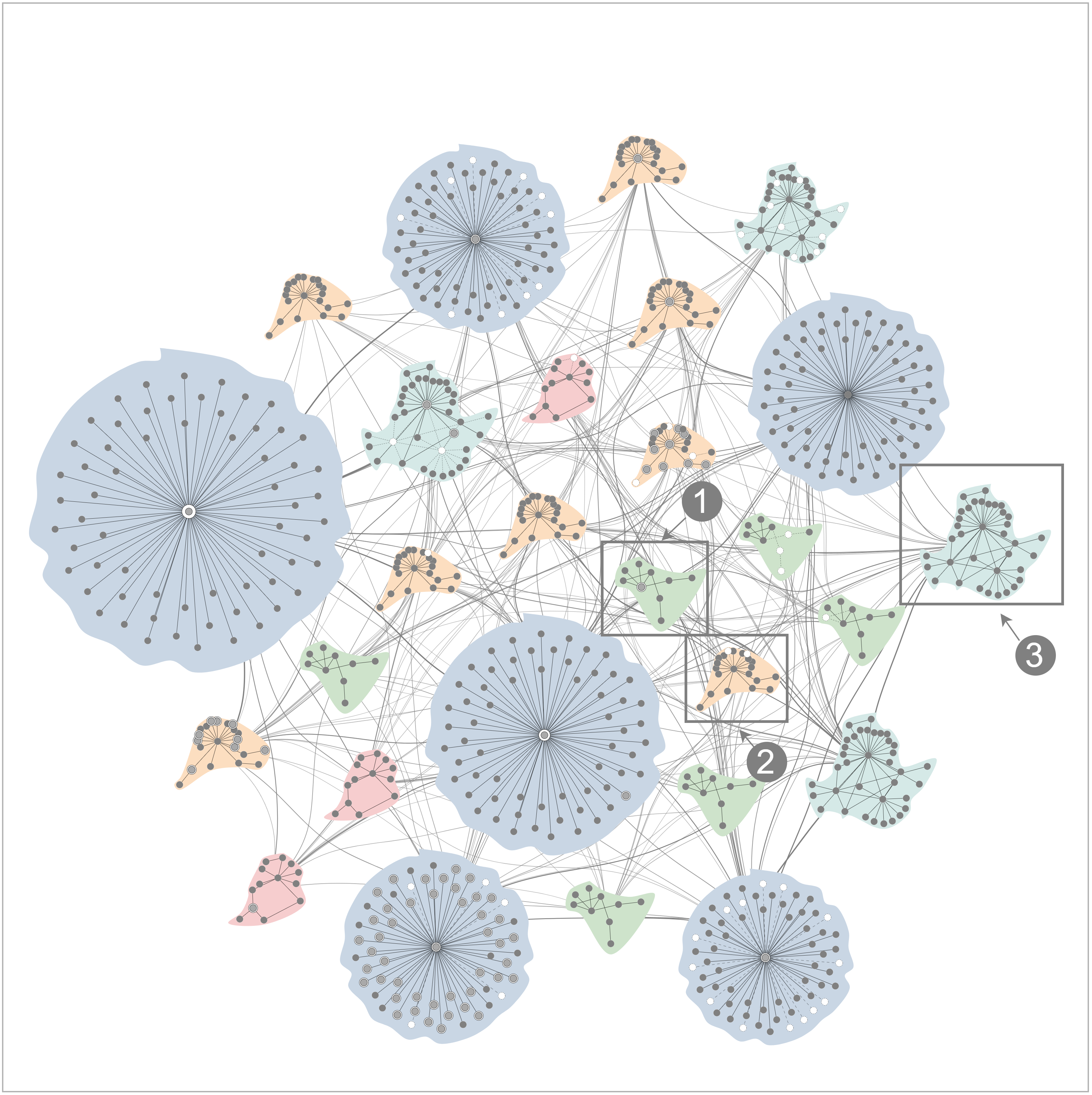} \\
(a) & (b)
 \end{tabular}
\begin{tabular}{cccccc}
\includegraphics[width=0.95in]{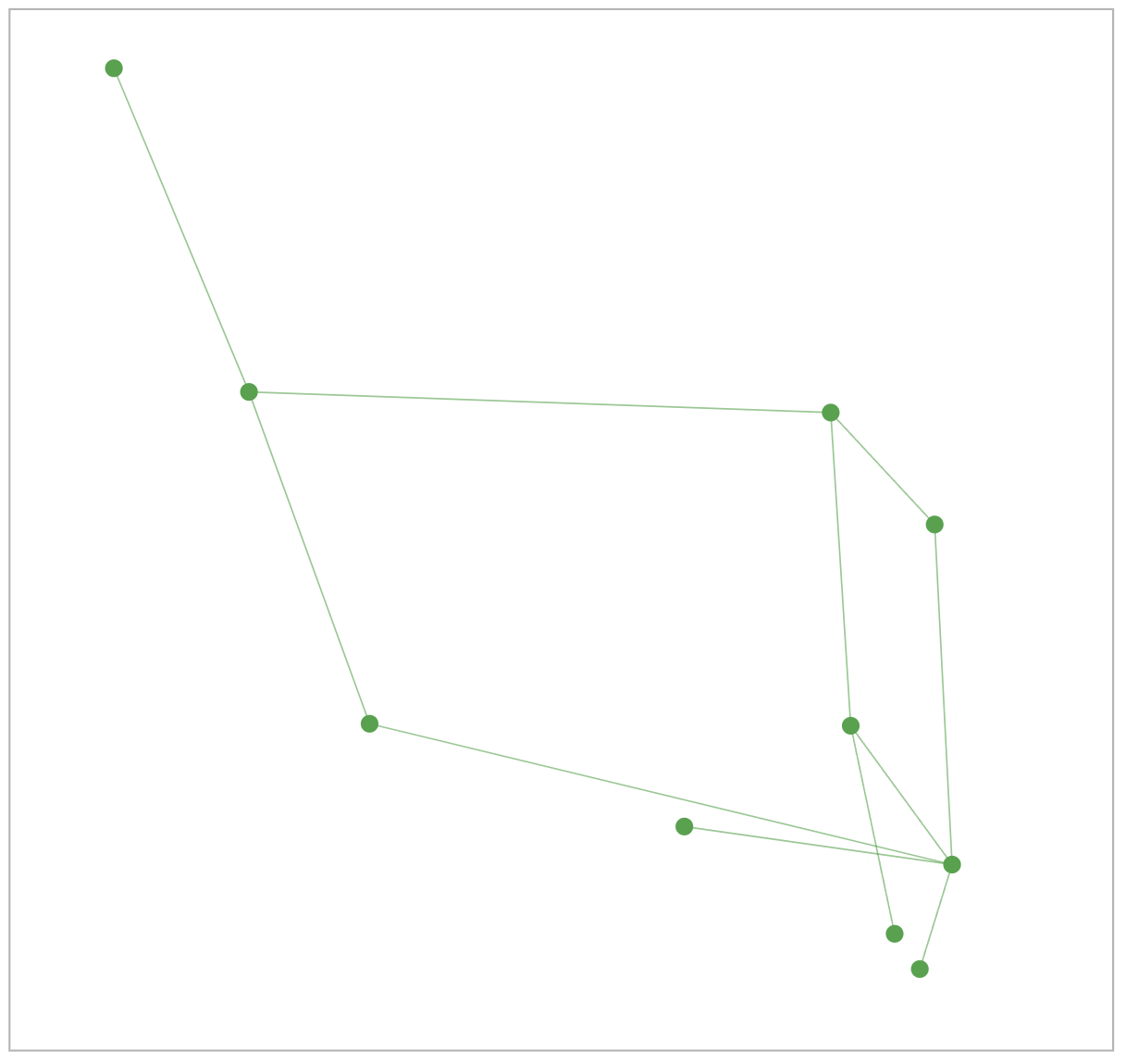} &  \includegraphics[width=0.95in]{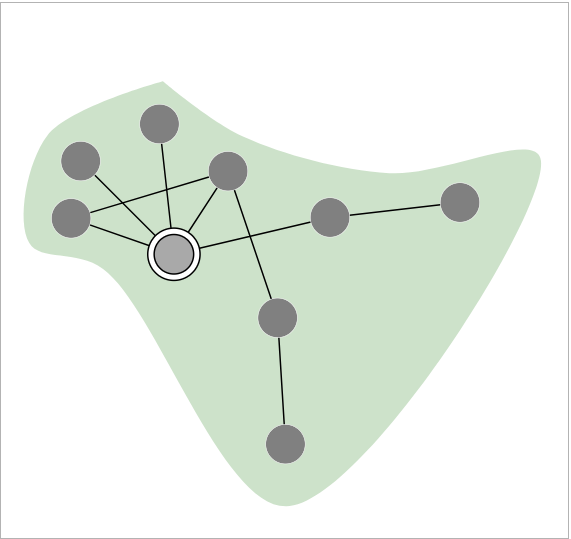} & \includegraphics[width=0.95in]{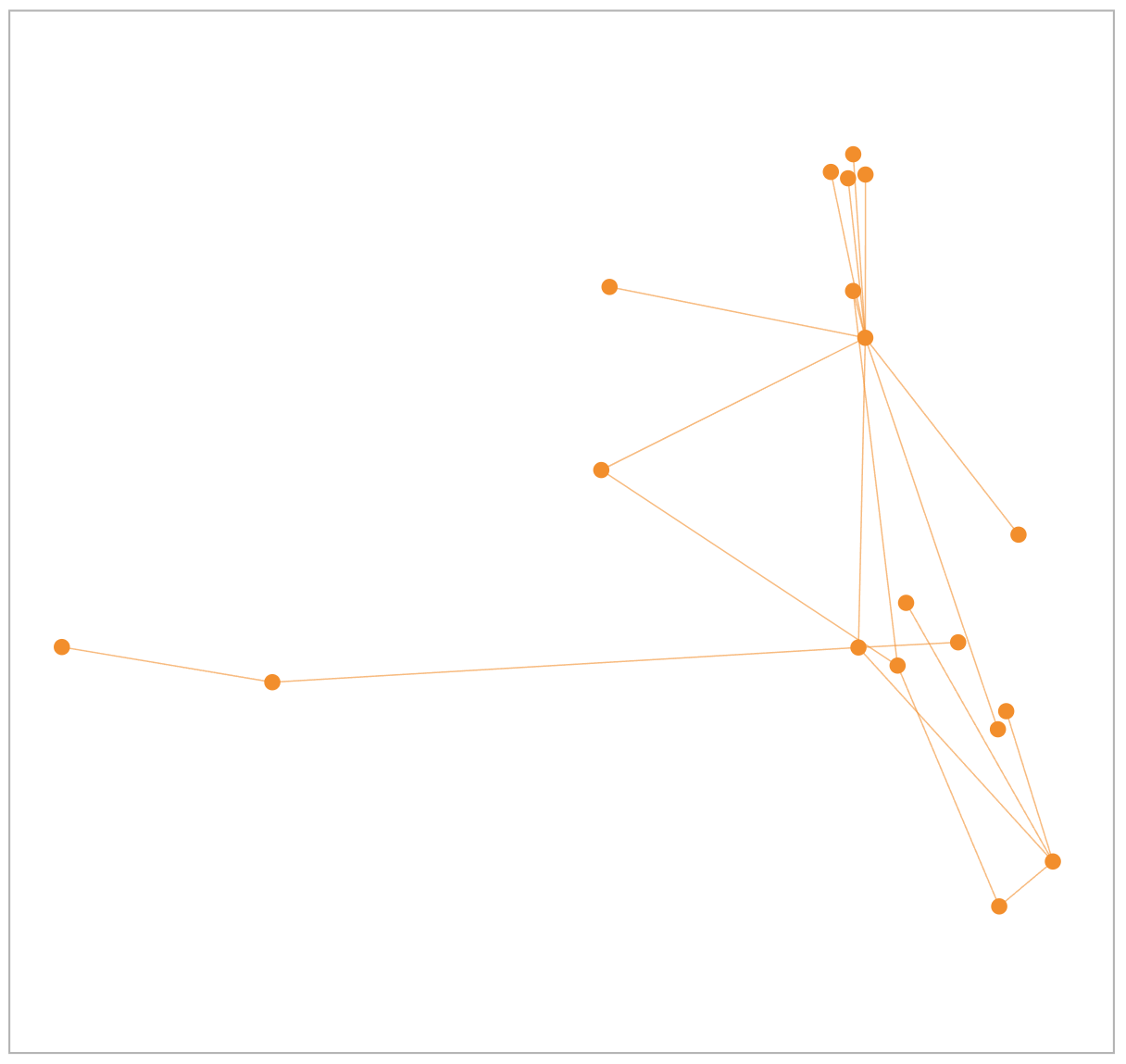} & \includegraphics[width=0.95in]{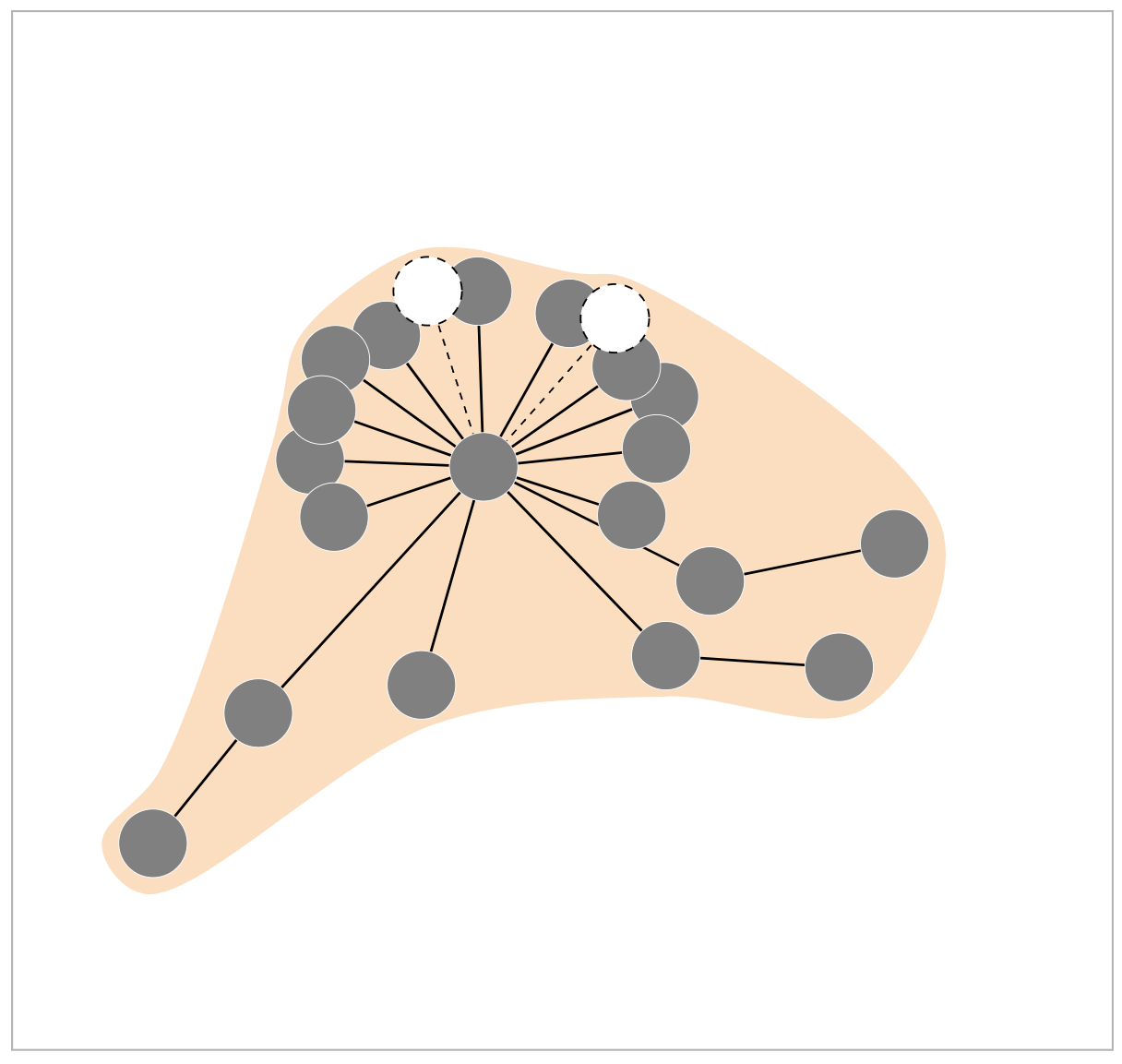} & \includegraphics[width=0.95in]{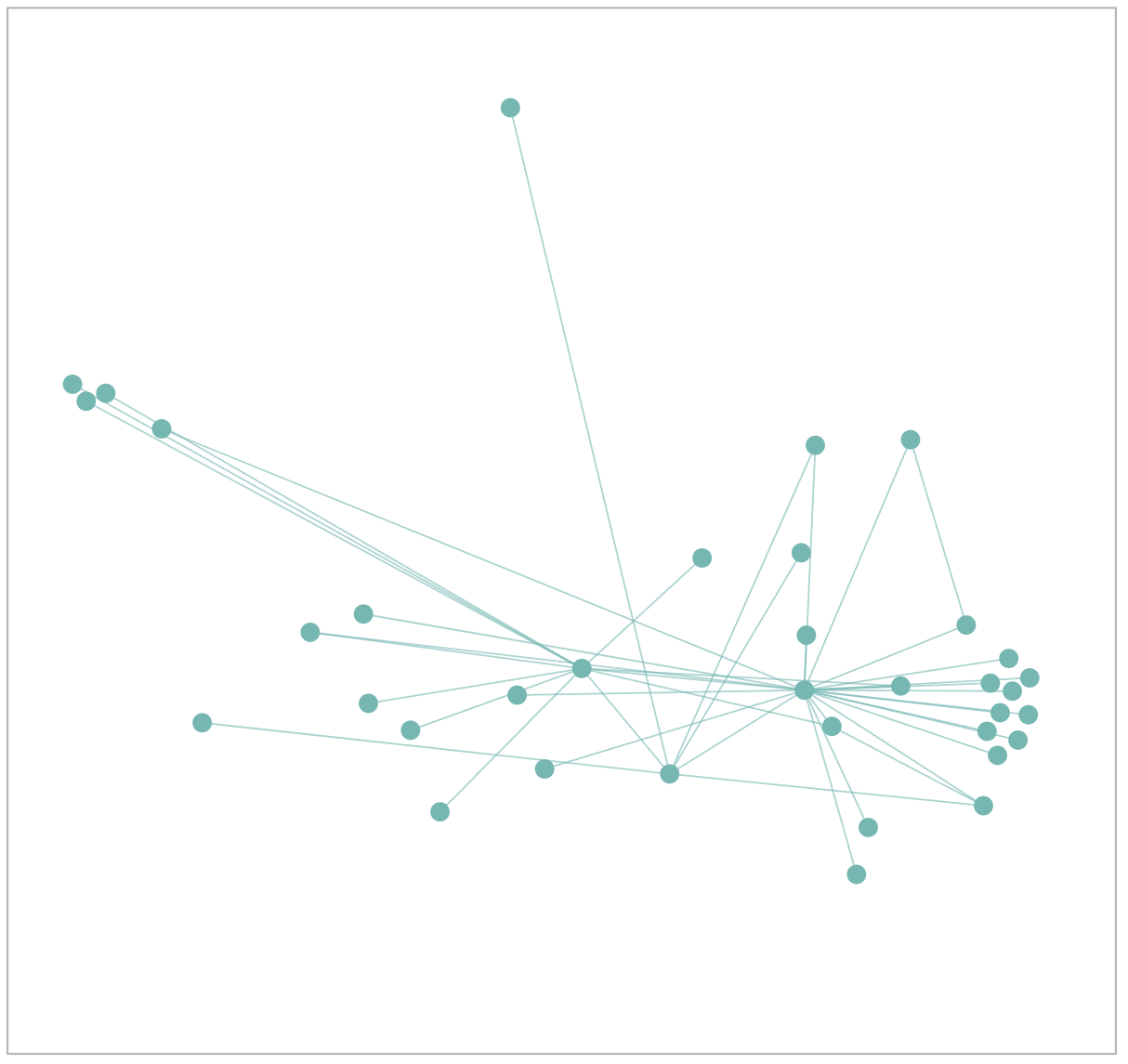} 
& \includegraphics[width=0.95in]{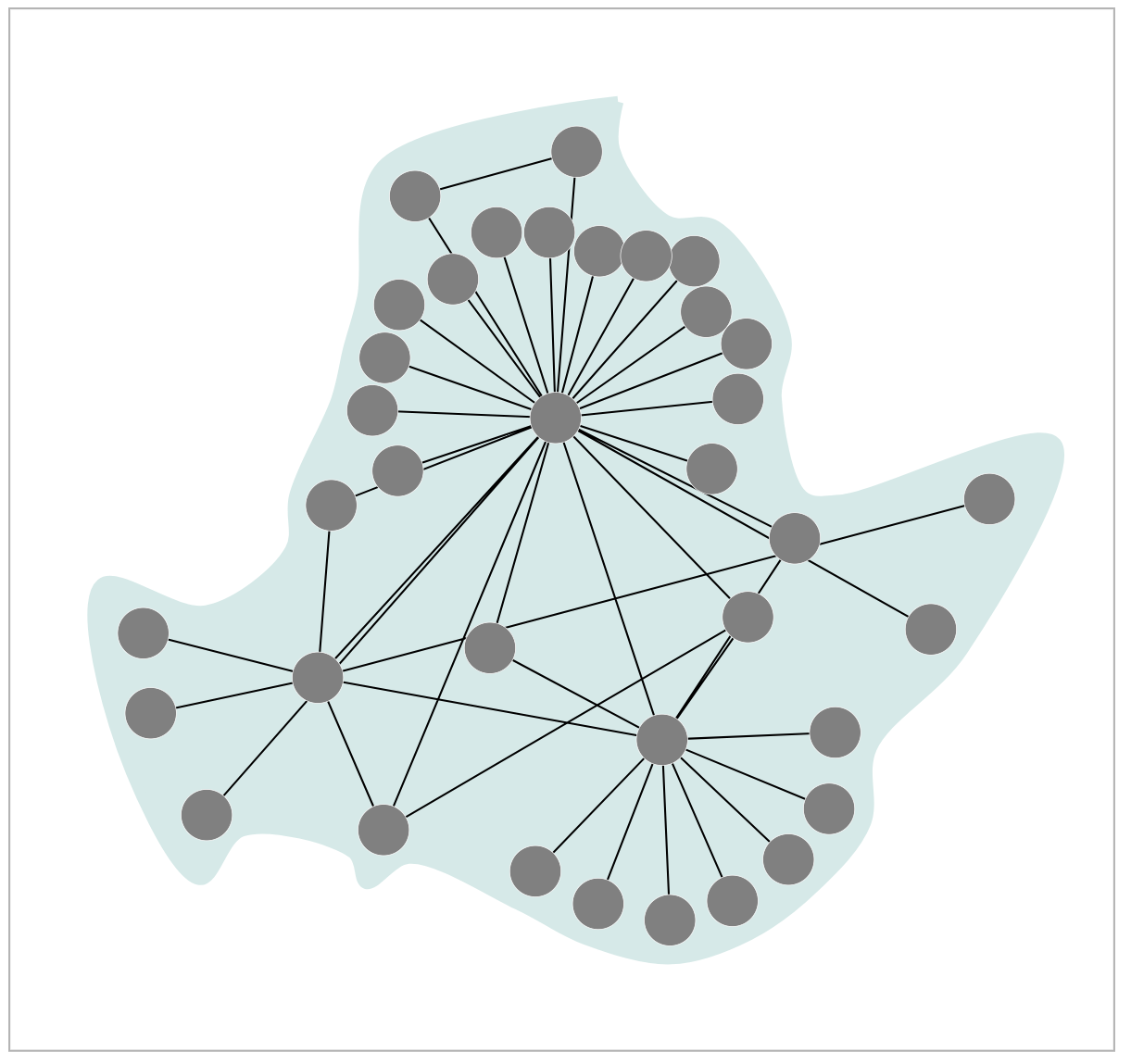}\\
(c) & (d) & (e) & (f) & (g) & (h)
\end{tabular}
  \caption{Cpan dataset~\cite{heymann2009cpan}: (a) the original graph with three communities in different colors; (b) the simplified graph with three motifs highlighted with gray boxes; (c), (e), and (g) separately displayed communities in different colors in (a); (d), (f), and (h) separately displayed motifs highlighted with gray boxes in (b). (c) and (d) represent the same community. (e) and (f) represent the same community. (g) and (h) represent the same community.
}
\label{fig:cpan_example}
\end{figure*}


\textbf{Difference-Aware Individual Subgraph Layout.}
For each subgraph cluster, we represent their similarities using the representative subgraph, but they also have differences. To illustrate their differences, we propose a difference-aware individual subgraph layout method, and an example is provided in Fig.~\ref{fig:diff}. For each non-representative subgraph (e.g., the graph in Fig.~\ref{fig:diff}(a)), we first compare it with the representative subgraph (e.g., the lower graph in Fig.~\ref{fig:diff}(b)) of its cluster using graph alignment to find their approximately identical parts (i.e., nodes with the same color in both graphs in Fig.~\ref{fig:diff}(c)). The unaligned parts (i.e., the gray nodes in Fig.~\ref{fig:diff}(c)) represent their differences. 

Since our adaptive motif is designed based on the representative subgraph, the differences should be illustrated based on the representative subgraph. Therefore, based on the representative subgraph, we identify three different types of nodes: 1) unaligned nodes that belong to the representative subgraph but do not have corresponding aligned nodes in the subgraph (e.g., the white node with a dashed boundary in Fig.~\ref{fig:diff}(d)); 2) aligned nodes connected with unaligned nodes that do not belong to the representative subgraph but have corresponding aligned nodes in the subgraph (e.g., the gray node with an outer ring in Fig.~\ref{fig:diff}(d)); and 3) aligned nodes that are not connected with any unaligned nodes not in the representative subgraph (e.g., the normal gray node in Fig.~\ref{fig:diff}(d)). For the first type of node, the edges connected to this node are also set as dashed lines. For the second type of node, the radius of the outer ring represents the number of unaligned nodes connected to this node. A larger radius indicates a larger number of connected unaligned nodes. Fig.~\ref{fig:encoding} illustrates all node encoding types in our difference-aware individual subgraph layout algorithm. Fig.~\ref{fig:encoding}(a) shows a white node with a dashed boundary, Fig.~\ref{fig:encoding}(b) displays a normal gray node, and Fig.~\ref{fig:encoding}(c)-(e) illustrate three different sizes of gray nodes with an outer ring that we employ. The radius of all the white node with a dashed boundary and gray nodes is the same, except for the outer rings of gray nodes. We use only three sizes of radius for the outer ring to represent the number of connected unaligned nodes. The radius of the outer ring in Fig.~\ref{fig:encoding}(c)-(e) corresponds to three levels of node numbers: small, medium, and large. This encoding scheme can help users roughly identify and distinguish different levels of node numbers, without the outer rings being too small to identify or too large to obscure nearby information. We can observe that using the layout of Fig.~\ref{fig:diff}(e) to represent Fig.~\ref{fig:diff}(a) makes it easier to identify the differences between a subgraph and its representative subgraph.

\subsection{Adaptive Motif Generation}

Our adaptive motifs are generated based on the subgraph Layout. That is, the motifs of representative subgraphs are generated based on the similarity-aware representative subgraph layout, while the motifs of other subgraphs are generated based on the difference-aware individual subgraph layout. The size of the motif indicates the number of nodes within the corresponding subgraph. To enhance the visibility of each motif, we add an outer contour and filled them with color. Different colors represent different subgraph clusters. The convex hull algorithm can be used to extract the contours of each subgraph~\cite{heer2005vizster}. However, due to the irregular shapes of each motif, some convex polygons do not wrap the motifs very tightly. Using concave polygons might be more suitable, allowing users to capture the distinctive features of the motifs.
Therefore, we adopt the alpha-shape algorithm~\cite{edelsbrunner1983shape}, which supports the extraction of concave polygon boundaries. For aesthetic purposes, we further utilize a polybuffer algorithm~\cite{polybuffer} to enlarge the outer contour in order to enclose nodes of certain sizes. As shown in Fig.~\ref{fig:workflow}(e), each motif is generated from the corresponding subgraphs in Fig.~\ref{fig:workflow}(c) and represented using our subgraph layout representation in Fig.~\ref{fig:workflow}(d). After generating all motifs, we then perform a force-directed layout to optimize the overall layout. 
Our motif design simplifies nodes and reduces edges in the graph. 
In our design, edges between nodes in the original graph are replaced by a single edge connecting motifs, even if multiple edges existed between nodes within those motifs in the original graph. We represent the number of edges connecting nodes within these two motifs in the original graph using the grayscale value of the edge. A darker grayscale value indicates a higher number of the original edges. Additionally, we use the edge bundling technique~\cite{holten2009force,d3-shape} to enhance the overall visual effect of our method. Fig.~\ref{fig:workflow}(f) illustrates the final simplified effect of Fig.~\ref{fig:workflow}(a). Our method is entirely automated and can adapt to simplify various datasets with different types of subgraphs.

\begin{figure}[t]
\centering
\begin{tabular}{c}
\includegraphics[width=0.8in]{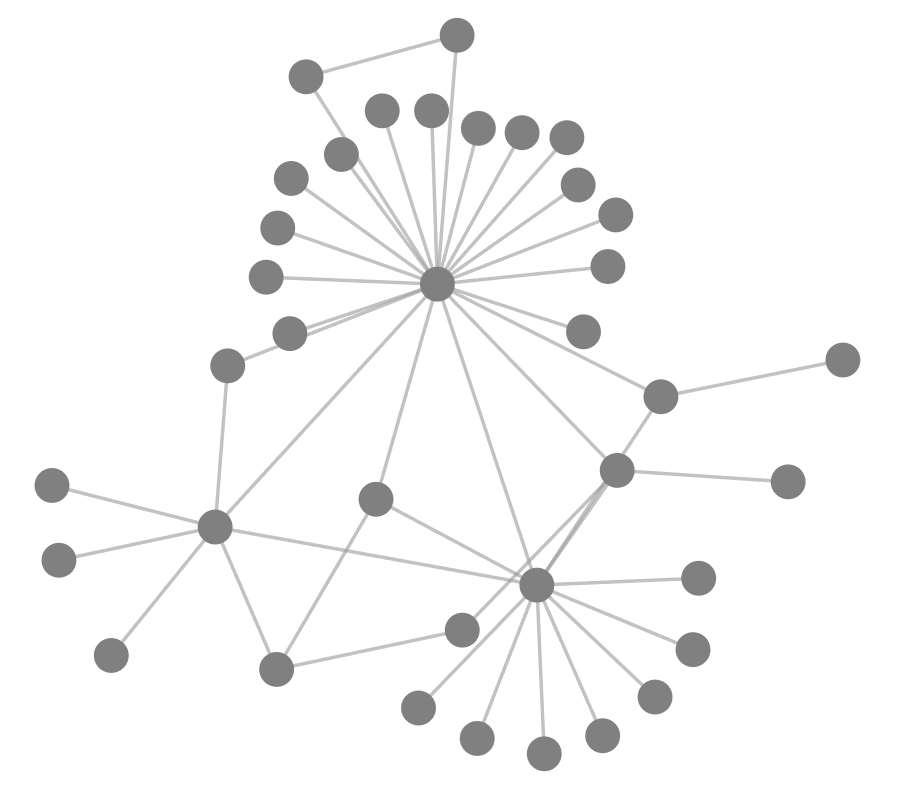}  \\
(a) \\
\end{tabular}
\begin{tabular}{ccc}
\includegraphics[width=.7in]{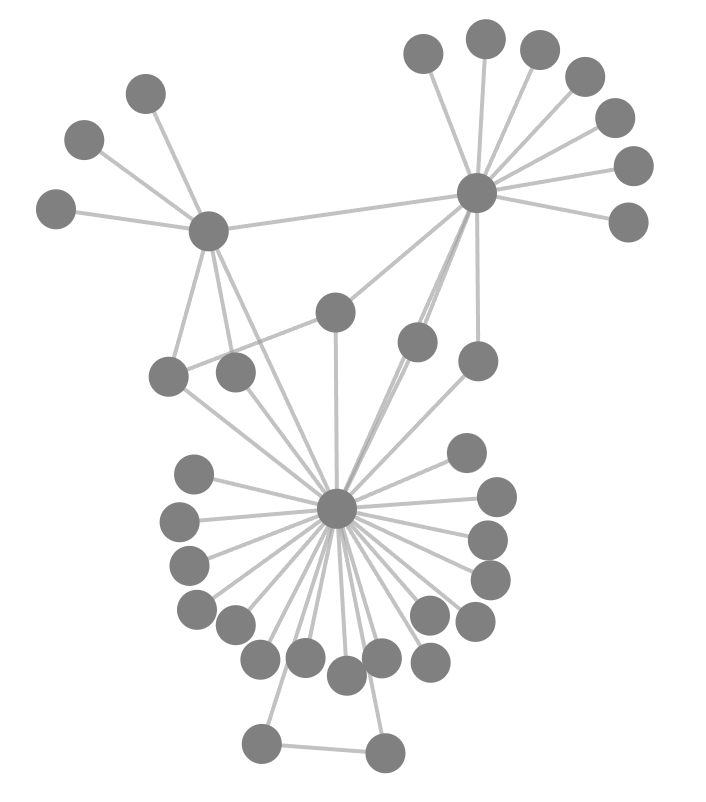} &  \includegraphics[width=0.9in]{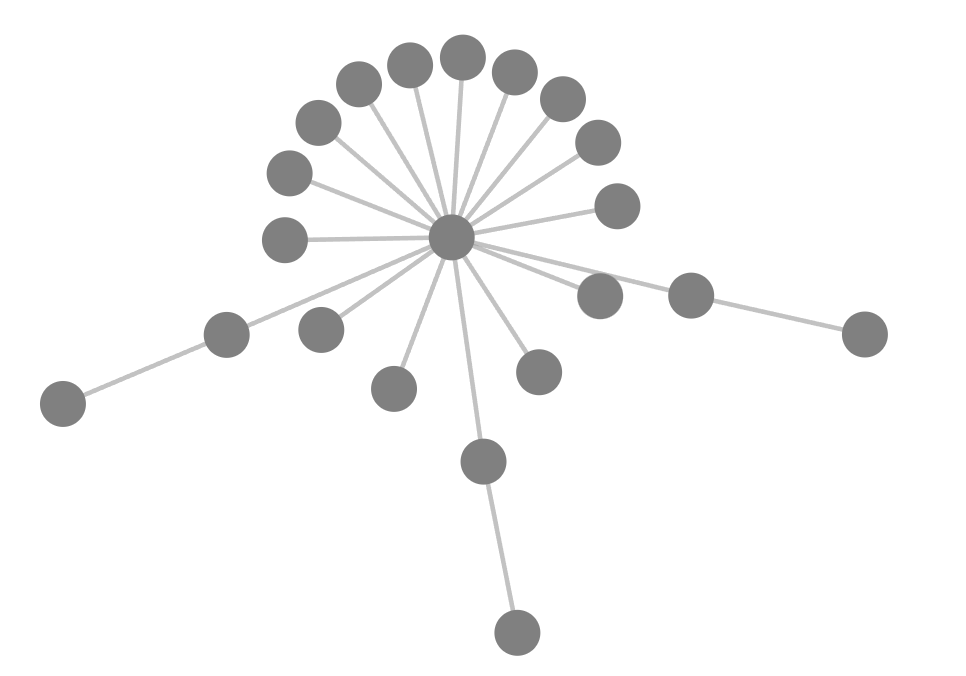} & \includegraphics[width=0.8in]{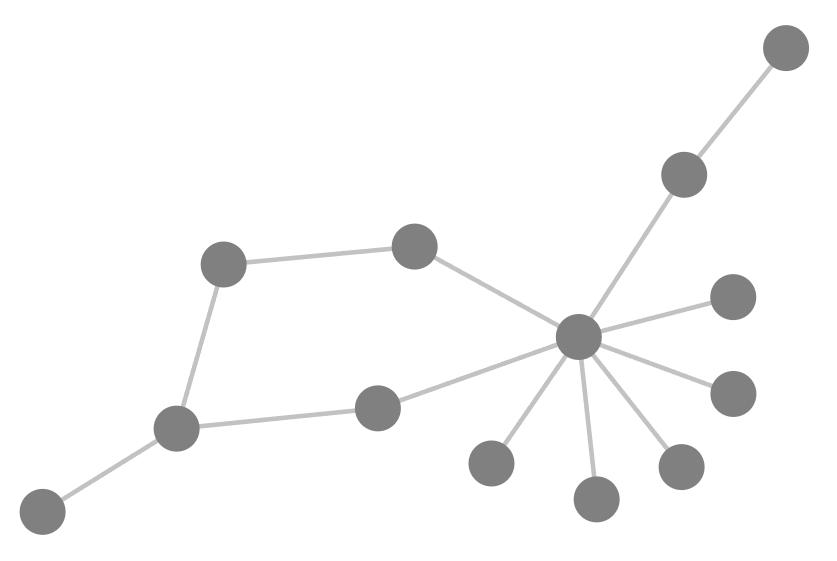} \\
(b) & (c) & (d)\\
\includegraphics[width=0.8in]{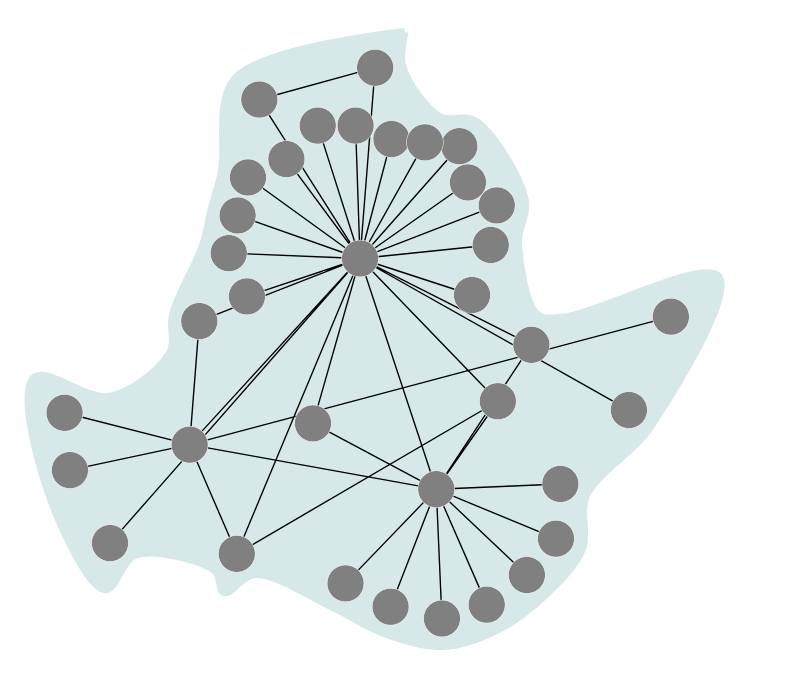} & \includegraphics[width=0.7in]{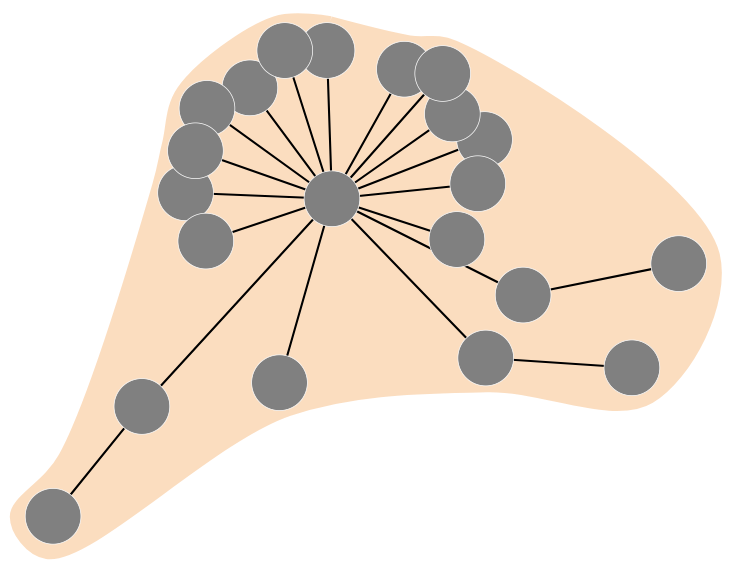} 
& \includegraphics[width=0.6in]{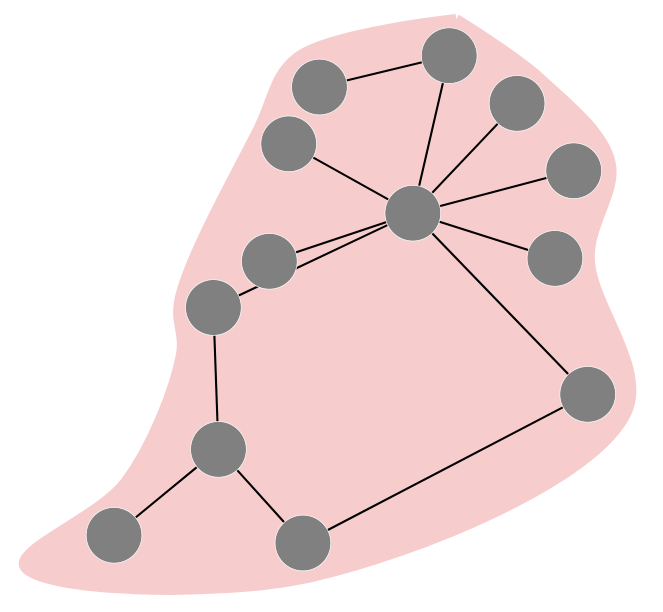}\\
(e) & (f) & (g)
\end{tabular}
 \caption{An example of the similarity-aware representative subgraph layout result: (a) a super-graph generated by (b)-(d) representative subgraphs, and (e)-(g) represent the corresponding layouts for (b)-(d) based on (a).
}
\label{fig:cpan_supergraph}
\end{figure}

\section{Case Study}

In this section, we apply our \tool{} method to three graphs of different scales and demonstrate the effectiveness of our approach. 

\subsection{Small Graph}

Firstly, we tested our method on a real small graph, which is the character network from \textit{Les Misérables}~\cite{knuth1993stanford} with 77 nodes and 254 edges. Due to its small scale, we can clearly see the overall graph structure in Fig.~\ref{fig:workflow}(a). The nine communities in the graph are labeled with different colors in Fig.~\ref{fig:workflow}(b). We can clearly see the overall graph structure and identify the six communities located on the periphery. However, the three communities in the middle (i.e., the orange, yellow, and cyan communities) are intertwined and difficult to distinguish, let alone compare them. If the communities are displayed independently (as shown in the left part of Fig.~\ref{fig:workflow}(c)), it is relatively easy to identify and compare. 
However, it becomes impossible to examine the community connections and the overall graph structure. 

In our results (see Fig.~\ref{fig:workflow}(f)), the communities are displayed as motifs, making them easy to be identified and compared. Furthermore, we can determine their scale and category information based on their size and color. Moreover, we can observe that the blue motif represents a star-shaped structure (i.e., a central node surrounded by many peripheral nodes), the yellow motif represents a clique structure (i.e., highly interconnected nodes), and the pink motif represents a grid structure (i.e., most nodes have balanced degrees). Motif ``1'' has a star-shaped structure, with the central node possibly being the main character. So, we checked the original information of the nodes and found that this is indeed the case. The central node represents Valjean, the protagonist of Les Misérables.
%
%
Additionally, the four yellow motifs represent communities of closely related characters. In the context of the novel, these four communities correspond to the characters involved in the Champmathieu case, the young revolutionaries, the characters associated with Fantine’s life, and those connected to Valjean’s growth and transformation. These close-knit communities play a pivotal role in the development of the novel’s narrative.

The differences between motifs of the same color (i.e., communities of the same type) can also be roughly discerned by the number of the white node with a dashed boundary and the gray node with an outer ring. In Fig.~\ref{fig:workflow}(f), the four blue labeled motifs have different sizes. Although motif "2" is the same size as the representative subgraph motif "1", it still has subtle structural differences compared to motif "1". This is because it contains one white node and one gray node with an outer ring. The other two slightly smaller motifs "3" and "4" contain multiple white nodes, indicating that they are simpler than the representative motif's star-shaped structure, resembling more like a small star dragging a long tail. These structural differences are difficult to discern in Fig.~\ref{fig:workflow}(b), let alone in Fig.~\ref{fig:workflow}(a). Although Fig.~\ref{fig:workflow}(c) can provide clear community similarities and differences, the community connections and the overall graph structure are missed.

However, the community connections are also clearly shown in our result (see Fig.~\ref{fig:workflow}(f)), since the number of edges is greatly reduced after graph simplification. For example, the blue motif in the bottom right corner and the orange motif on the rightmost are connected by a dark gray edge label with ``5'', indicating that there are many edges connecting nodes between these two motifs. However, specific edge information between nodes is lost. This is an inevitable consequence of information loss caused by graph simplification.

Therefore, our method can help users identify communities and compare their similarities and differences, while also allowing them to examine the community connections and the overall graph structure. This cannot be achieved solely with Fig.~\ref{fig:workflow}(b) or (c). This example demonstrates that our method works well for small graphs.

\subsection{Medium-Sized Graph}

Secondly, we tested our method on a medium-sized graph named Cpan~\cite{heymann2009cpan}, which is a collaboration network with 839 nodes and 2,127 edges. This graph depicts the relationships between developers using the same Perl modules. Fig.~\ref{fig:Cpan-teaser}(a) shows the original graph with the force-directed layout, while Fig.~\ref{fig:Cpan-teaser}(b) illustrates our simplified result. In Fig.~\ref{fig:Cpan-teaser}(a), only communities with big star-shaped structure can be clearly observed, one of them is highlighted with gray box labeled as ``1''. Communities with other structures, such as clique structure, grid structure, or more general structures, are difficult to identify due to the high density of nodes and edges.

In Fig.~\ref{fig:Cpan-teaser}(b), communities with both big star-shaped structure and other structures can be easily identified. The community labeled as ``1'' and ``2'' in the gray box are the same one. We can observe that our motif effectively captures the star-shaped structure of this community. Many structures that are not distinguishable in Fig.~\ref{fig:Cpan-teaser}(a) can be clearly identified in Fig.~\ref{fig:Cpan-teaser}(b). For example, the community ``3'' and ``4'' in the gray box are the same one. The nodes of community ``3'' are highlighted in red in Fig.~\ref{fig:Cpan-teaser}(a), and it can be observed that these nodes are relatively scattered, making community ``3'' difficult to identify. However, our motif for community ``4'' effectively captures the structure of this community and is easily recognizable.


In Fig.~\ref{fig:cpan_example}(b), users can observe that all the communities are divided into five categories based on the motif colors. Three motifs from different community categories are highlighted with gray boxes and numbered. Their corresponding communities are difficult to identify in Fig.~\ref{fig:cpan_example}(a), even though we have also highlighted them with corresponding colors. The community corresponding to the green motif labeled as ``1'' in Fig.~\ref{fig:cpan_example}(b) is also marked in green in Fig.~\ref{fig:cpan_example}(a).  The same applies to the other two numbered motifs. In Fig.~\ref{fig:cpan_example}(a), the three communities colored in yellow and green are intertwined, making them difficult to identify and even harder to compare. However, they are easily identifiable and comparable in our results shown in Fig.~\ref{fig:cpan_example}(b), due to the use of different motifs. Fig.~\ref{fig:cpan_example}(c) and (d) represent the same community. Fig.~\ref{fig:cpan_example}(e) and (f) represent the same community. Fig.~\ref{fig:cpan_example}(g) and (h)
represent the same community. Through comparison, it can be observed that our motif effectively represents the community scattered in the original graph, making them clear and easy to identify. 

Furthermore, we verified the effectiveness of our similarity-aware representative subgraph layout result. By examining the intermediate results of our program, we found that the five representative subgraphs of this graph have been divided into two classes: one containing two subgraphs and the other containing three. The super-graph generated by the class with three representative subgraphs is shown in Fig.~\ref{fig:cpan_supergraph}(a), and the three representative subgraphs before and after super-graph alignment are shown in Fig.~\ref{fig:cpan_supergraph}(b)-(g). We found that the alignment effect of the super-graph is quite evident. In Fig.~\ref{fig:cpan_supergraph}(b)-(d), there is a prominent star-shaped node in each, representing nodes with high similarity. However, in Fig.~\ref{fig:cpan_supergraph}(b), this node is located below; in Fig.~\ref{fig:cpan_supergraph}(c), it is located above; and in Fig.~\ref{fig:cpan_supergraph}(d), it is located to the right, resulting in inconsistent positioning, which makes it difficult to compare them. In contrast, in our results in Fig.~\ref{fig:cpan_supergraph}(e)-(g), these prominent star-shaped nodes are all positioned at the top of the graph, making them easy to compare. Therefore, our similarity-aware representative subgraph layout algorithm can effectively demonstrate the similarities and differences between communities.


\begin{figure}[ht]
\centering
 \includegraphics[width=2.in]{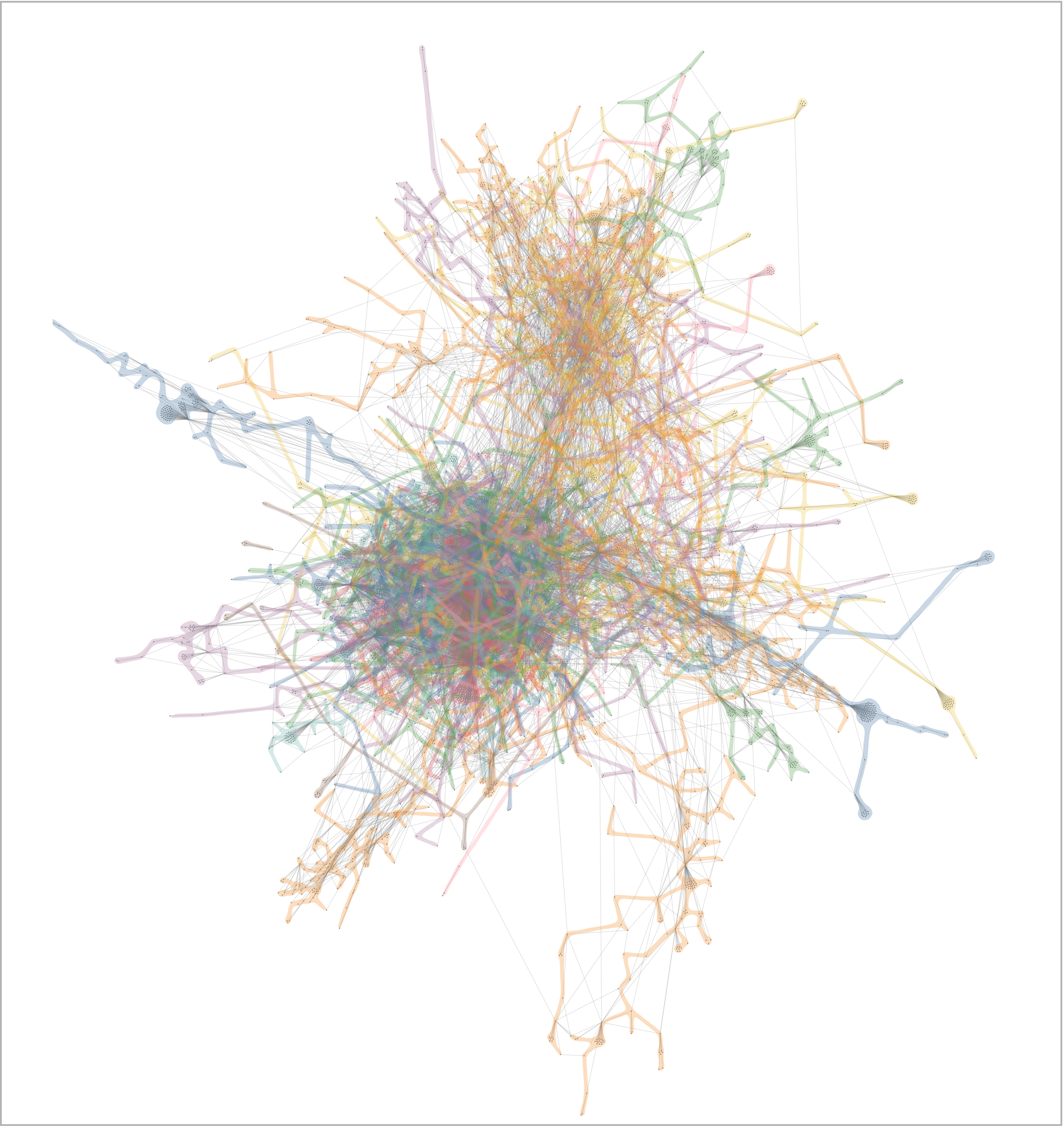} 
\caption{Using both color and Bubble Sets~\cite{collins2009bubble} to show communities in the AS-733 dataset.
}
  \label{fig:AS-733-bubbleset}
\end{figure}

\begin{figure*}[ht]
\centering
\begin{tabular}{cc}
 \includegraphics[width=2.7in]{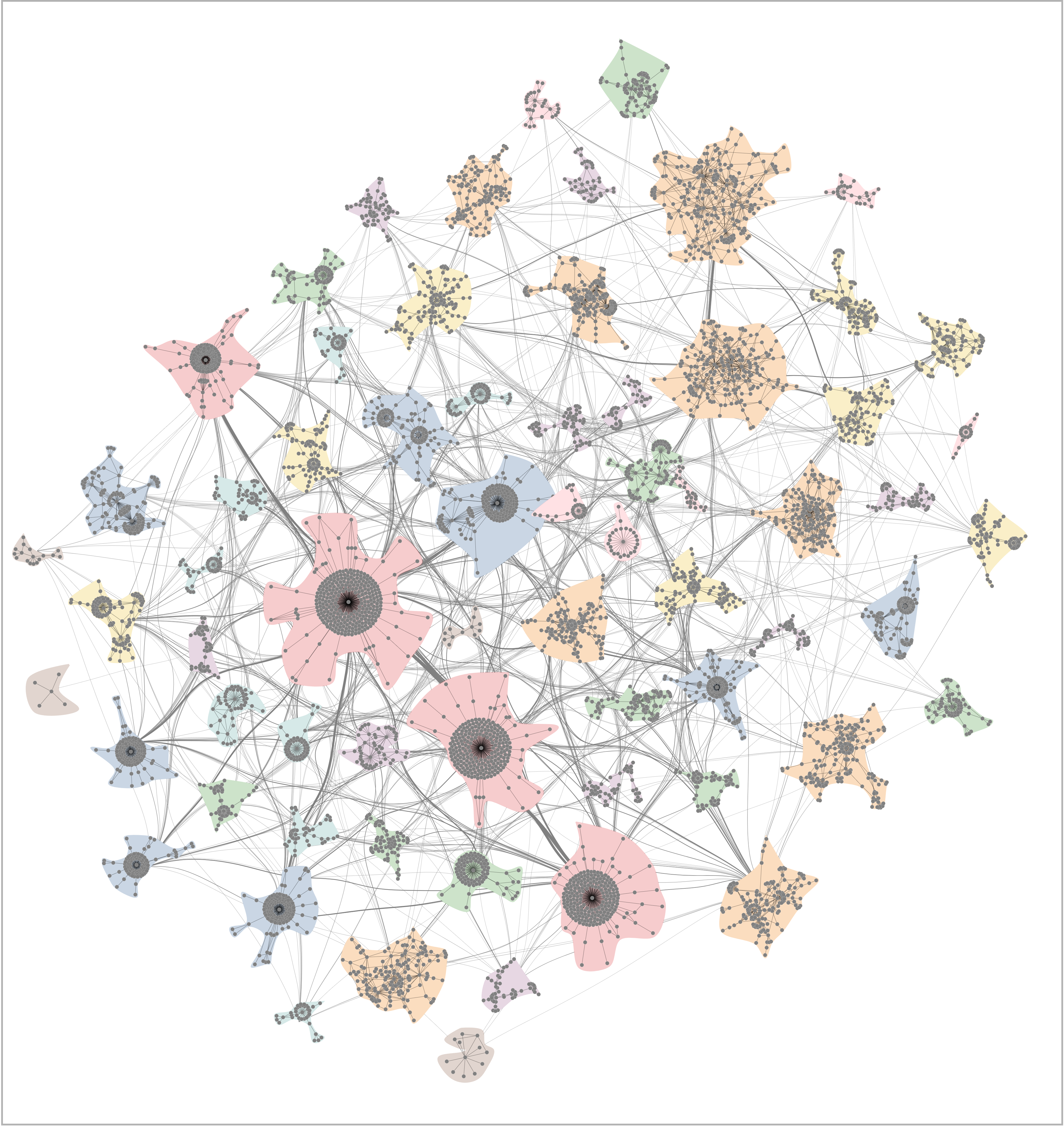} &  
 \includegraphics[width=2.7in]{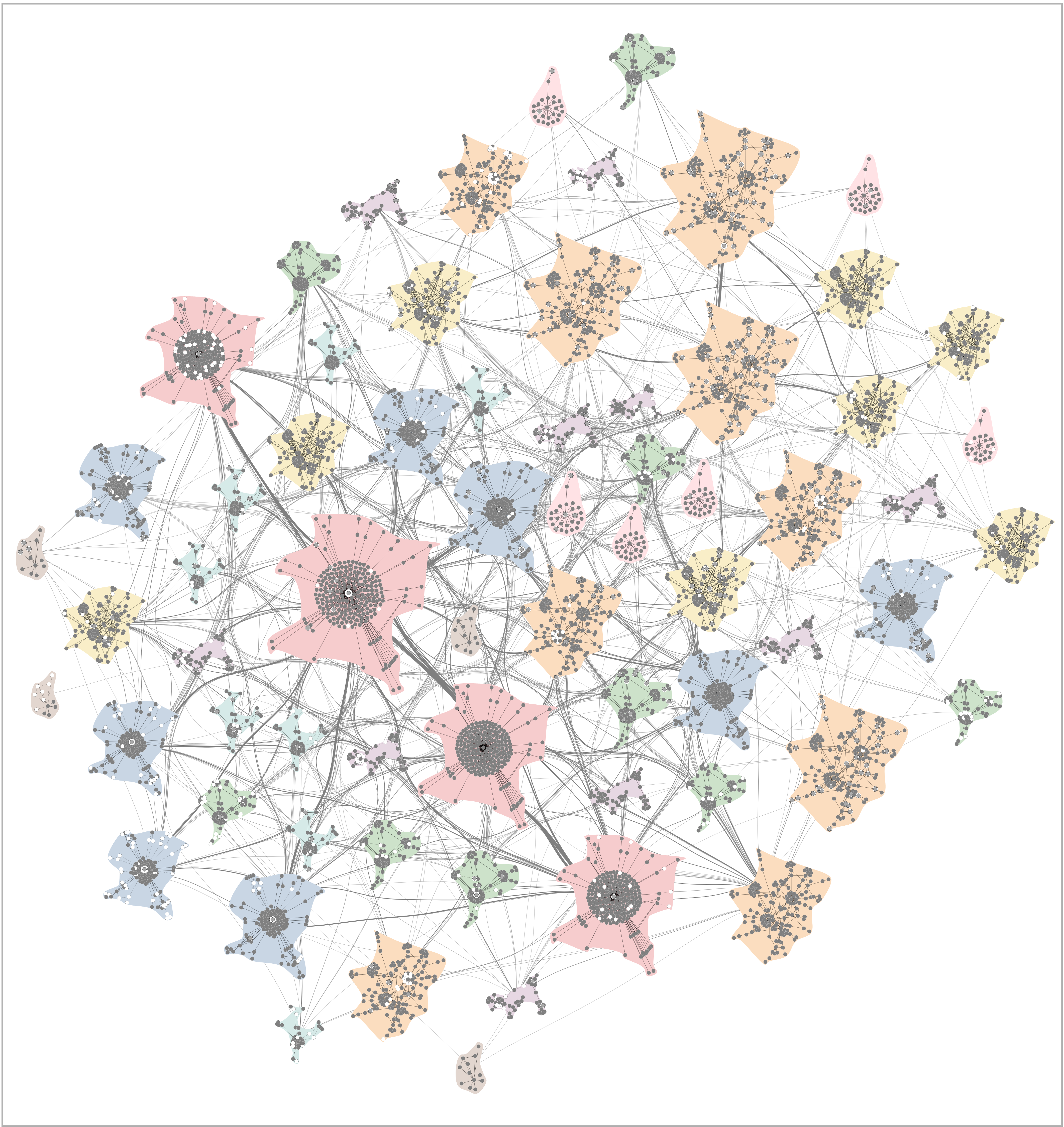}\\
(a) & (b)
\end{tabular}
\caption{AS-733 dataset: (a) the \textit{Primitive AdaMotif} result that executes our subgraph partitioning and hierarchical clustering but does not execute our subgraph layout; (b) our \tool{} result.}
  \label{fig:AS-733}
\end{figure*}


\subsection{Large Graph}

Finally, we tested our method on a large graph named AS-733~\cite{leskovec2005graphs}, which is an autonomous systems network on the Internet with 6,474 nodes and 13,895 edges. From the previous case study about Cpan, we observed that for medium-sized graphs, the community information becomes difficult to recognize from the original graph visualization, even when using colors to encode the communities (see the colored communities in Fig.~\ref{fig:Cpan-teaser}(a) and \ref{fig:cpan_example}(a)). Therefore, for the AS-733 graph, which is much larger than Cpan, we tried to represent different communities using both color and Bubble Sets~\cite{collins2009bubble} in the original graph visualization (see Fig.~\ref{fig:AS-733-bubbleset}). As shown in Fig.~\ref{fig:AS-733-bubbleset}, the scale of the graph has indeed become quite large. Although using Bubble Sets can better encode communities with scattered nodes, it is still difficult to recognize any community in areas with too many overlapping communities (e.g., the central area in Fig.~\ref{fig:AS-733-bubbleset}) due to the overlapping of many Bubble Sets. Therefore, it is necessary to represent the nodes of each community in a consolidated manner (e.g., our motif shapes) and avoid overlap.

Next, we implemented a simplified version of \tool{}, called \textit{Primitive AdaMotif}.
\textit{Primitive AdaMotif} executes our subgraph partitioning and hierarchical clustering, but does not execute our subgraph layout. It also uses our outer contour and colors to represent each community. The size of the outer contour indicates the number of nodes in the community. To facilitate the comparison between \textit{Primitive AdaMotif} and \tool{}, we used the same encoding scheme to represent the edges between motifs and applied the edge bundling technique. Their results applied to the AS-733 graph are shown in Fig.~\ref{fig:AS-733}(a) and (b), respectively. In both results, the graph is simplified into multiple structures, significantly reducing the scale of the graph and enhancing the utilization of screen space. Additionally, we can quickly identify four big star-shaped communities (i.e., the four large pink communities). Despite the large number of nodes, we can still identify multiple communities and their categories via the colors from the results of both \textit{Primitive AdaMotif} and \tool{}, as shown in Fig.~\ref{fig:AS-733}(a) and (b).

However, it is hard to compare communities to identify their similarities and differences by using \textit{Primitive AdaMotif}. On the contrary, \tool{} shows clear advantages.
Firstly, for comparing different types of communities (i.e., communities of different colors), as shown in Fig.~\ref{fig:AS-733}(b), it is easy to see that the pink and blue motifs are both main star-shaped communities, while the green motif contains multiple star-shaped communities. Given the diverse community shapes in Fig.~\ref{fig:AS-733}(a), it is not easy to draw such conclusions by visual inspection alone. Secondly, for comparing communities of the same type (i.e., communities of the same color), it is relatively easy to distinguish the differences among the four large pink star-shaped motifs located in the middle of Fig.~\ref{fig:AS-733}(b). This is because, based on the graph alignment result, white nodes (indicating that this community actually does not have this node compared to the representative subgraph) and gray nodes with an outer ring (indicating that there are still some nodes in this community that are not displayed) are marked, which facilitates visual comparison. 
This last example demonstrates that our method performs well on large graphs as well and shows its advantages over the original graph visulaization and the result of \textit{Primitive AdaMotif}.

We implemented our~\tool{} method on a desktop with the Windows 10 operating system, Intel Core i3-12100F 3.3GHz CPUs, 
NVIDIA RTX 2060 12GB graphics card, 
and 16GB of memory. The computation times of our method for datasets used in Fig.~\ref{fig:workflow}, Fig.~\ref{fig:cpan_example}, and Fig.~\ref{fig:AS-733} are 15.12s, 21.9s, and 41.46s respectively. 

\section{Expert Interview}
\revise{
To further assess the effectiveness of \tool{},
we presented the AS-733 graph to a computer network expert with over 15 years of experience 
in High-Performance Routing and Internet Architecture. 
He found \tool{} beneficial for research, as it classifies Autonomous System (AS) communities and provides a global view of Internet topology. 
Our method effectively highlights similarities and differences between communities, aiding in identifying critical parts of the Internet, optimizing performance, and improving security.
By analyzing the similarities and differences between AS communities, the security of each AS can be assessed. 
For example, it is critical to monitor and analyze
the security incidents of computer networks, which can exist in different communities~\cite{raynor2022state}.
With \tool{}, the network security practitioners and researchers can investigate the similarities and differences of communities that have experienced security incidents, enabling them to identify and respond to security threats.}
He also noticed that our motifs lose some information about the original nodes and edges. He suggested adding more interactive features to help them view such information simultaneously. 
Overall, our method is very promising, and he
is eager to apply \tool{} to  visualize other large graphs of computer networks he needs to analyze.

\begin{table}[t]
\renewcommand{\arraystretch}{1.0} 
\centering
\caption{The accuracy and time taken for tasks between the original graph and the~\tool{}. The tested datasets are classified into three scales, and the tasks are divided into three categories. }
\begin{tabular}{c@{\hspace{0.3cm}}c@{\hspace{0.3cm}}c@{\hspace{0.3cm}}c@{\hspace{0.3cm}}c@{\hspace{0.3cm}}c}
\hline
\multirow{2}{*}{Scale}                                                           & \multirow{2}{*}{Task} & \multicolumn{2}{c}{Accuracy}                                              & \multicolumn{2}{c}{Time(s)}                                                \\
                                                                                 &                       & \begin{tabular}[c]{@{}c@{}}Original\\ graph\end{tabular} & \tool{}       & \begin{tabular}[c]{@{}c@{}}Original\\ graph\end{tabular} & \tool{}         \\ \hline
\multirow{3}{*}{\begin{tabular}[c]{@{}c@{}}Small\\ graph\end{tabular}}           & Similarity            & 0.433                                                    & \textbf{0.817} & 30.38                                                   & \textbf{28.403} \\
                                                                                 & Difference            & 0.867                                           & 0.867          & 29.305                                                   & \textbf{21.360} \\
                                                                                 & Graph size            & 0.517                                                    & \textbf{0.750} & \textbf{23.790}                                          & 25.055 \\ \hline
\multirow{3}{*}{\begin{tabular}[c]{@{}c@{}}Medium\\ -sized\\ graph\end{tabular}} & Similarity            & 0.350                                                    & \textbf{0.817} & 46.019                                                   & \textbf{30.163} \\
                                                                                 & Difference            & \textbf{0.683 }                                                   & 0.650 & 29.445                                                   & \textbf{27.218} \\
                                                                                 & Graph size            & 0.383                                                    & \textbf{0.467} & 49.518                                          & \textbf{47.075}          \\ \hline
\multirow{3}{*}{\begin{tabular}[c]{@{}c@{}}Large\\ graph\end{tabular}}           & Similarity            & 0.250                                                    & \textbf{0.900} & 31.775                                                   & \textbf{20.500} \\
                                                                                 & Difference            & 0.600                                                    & \textbf{0.733} & 19.763                                                   & \textbf{14.802} \\
                                                                                 & Graph size            & 0.183                                                    & \textbf{0.317} & \textbf{27.005}                                          & 32.403          \\ \hline
\end{tabular}
\label{tab:result}
\end{table}

\begin{table}[h]
\renewcommand{\arraystretch}{1.0} 
\centering
\caption{The results of the normal distribution test for the objective questions ($*, p < 0.05$), where * indicates that the data significantly deviates from a normal distribution.}
\begin{tabular}{ccrrrr}
\hline
Task                        & Method         &$W_{acc}$ & $p_{acc}$ & $W_{time}$ & $p_{time}$  \\ \hline
\multirow{2}{*}{Similarity} & Original graph &0.938& 0.641 & 0.565& <0.001*                                  \\
                            & ~\tool{}       &0.919& 0.504 & 0.793& <0.001*                                  \\ \hline
\multirow{2}{*}{Difference} & Original graph &0.908& 0.425 & 0.697& <0.001*                                  \\
                            & ~\tool{}        &0.975& 0.924 & 0.671& <0.001*                                 \\ \hline
\multirow{2}{*}{Graph size} & Original graph &0.884& 0.285 & 0.585& <0.001*                                  \\
                            & ~\tool{}        &0.863& 0.198 & 0.775& <0.001*                                 \\ \hline
\end{tabular}
\label{tab:Normal1}
\end{table}

\begin{table}[h]
\renewcommand{\arraystretch}{1.0} 
\centering
\caption{The results of the normal distribution test for the subjective questions ($*, p < 0.05$), where * indicates that the data significantly deviates from a normal distribution.}
\begin{tabular}{ccrr}
\hline
Task                                                                          & Method         & $W$ & $p$        
\\ \hline
\multirow{2}{*}{\begin{tabular}[c]{@{}c@{}}Overall \\ structure\end{tabular}} & Original graph &0.952& 0.019 *                  \\
                                                                              & ~\tool{}        &0.794& 0.000 *                   \\ \hline
\multirow{2}{*}{\begin{tabular}[c]{@{}c@{}}Local \\ structure\end{tabular}}   & Original graph &0.886& 0.004 *                   \\
                                                                              & ~\tool{}        &0.845& 0.000 *                  \\ \hline
\end{tabular}
\label{tab:Normal2}
\end{table}

\begin{figure}[t]
\centering
\begin{tabular}{cc}
 \includegraphics[width=1.6in]{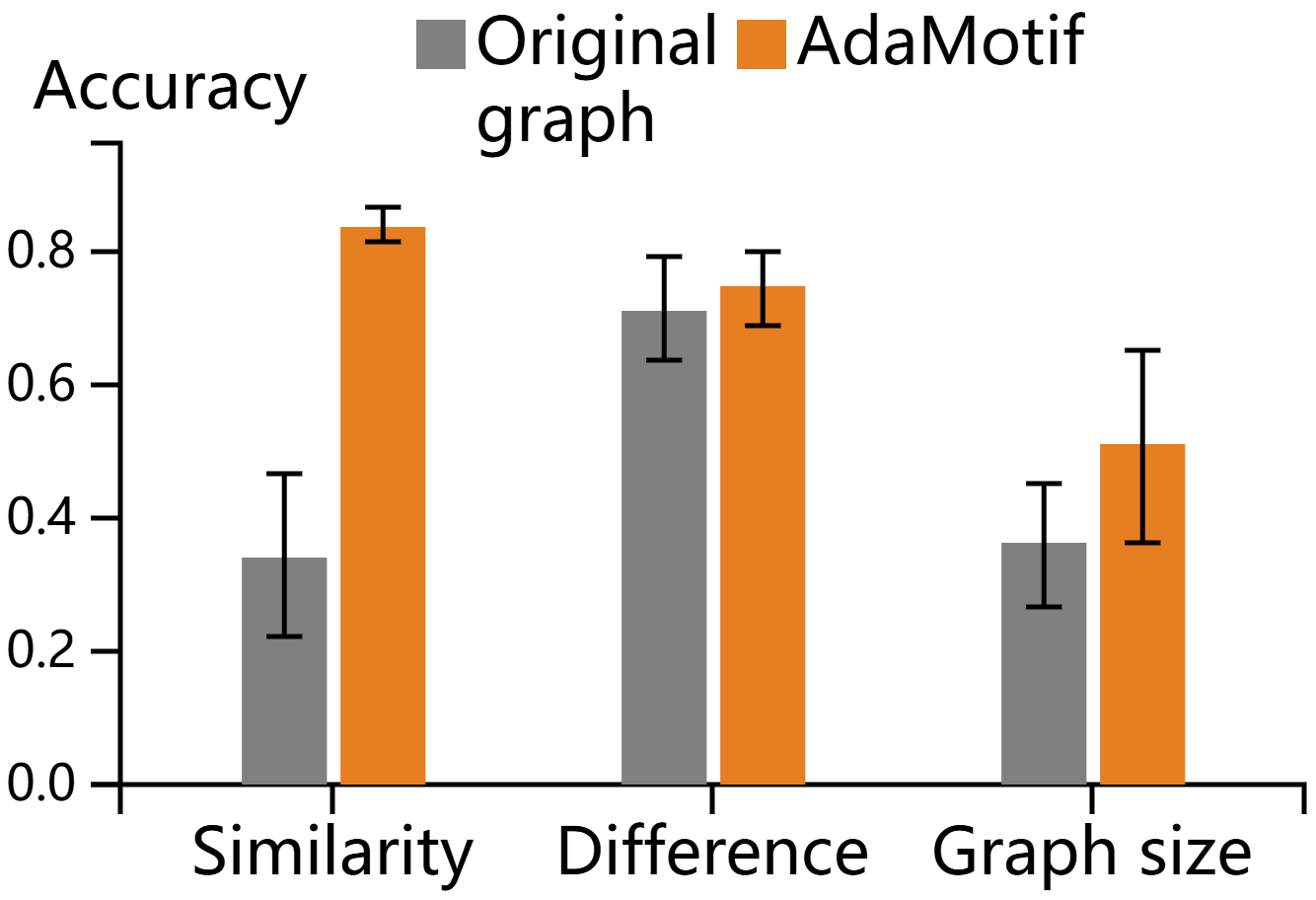} &  \includegraphics[width=1.6in]{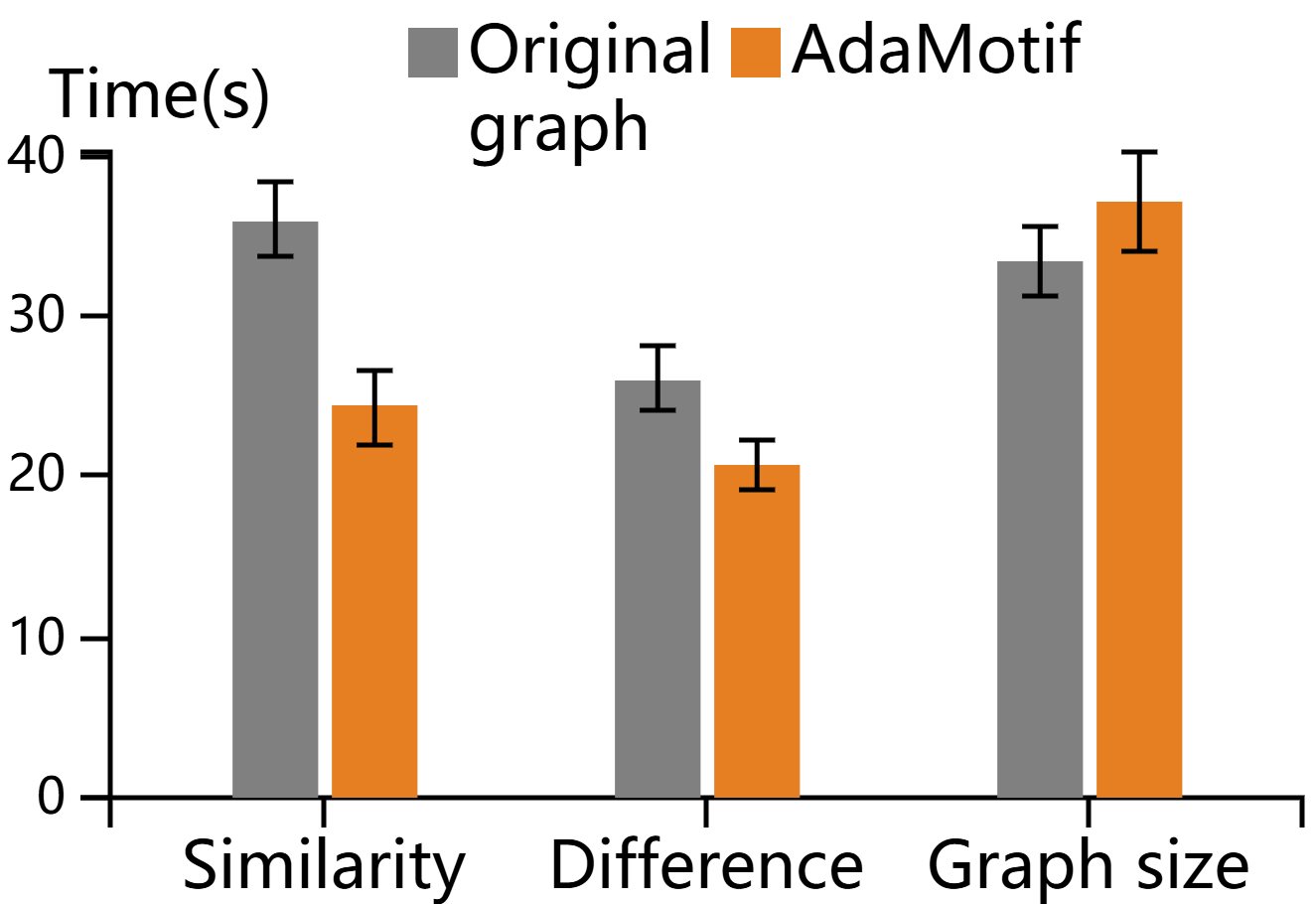}\\
(a) & (b)
\end{tabular}
  \caption{Means and standard errors of the Original graph and~\tool{} on \textit{similarity}, \textit{difference} and \textit{graph size} ($*, p < 0.05$).
}
  \label{fig:task}
\end{figure}

\begin{figure}[h]
\centering
 \includegraphics[width=1.58in]{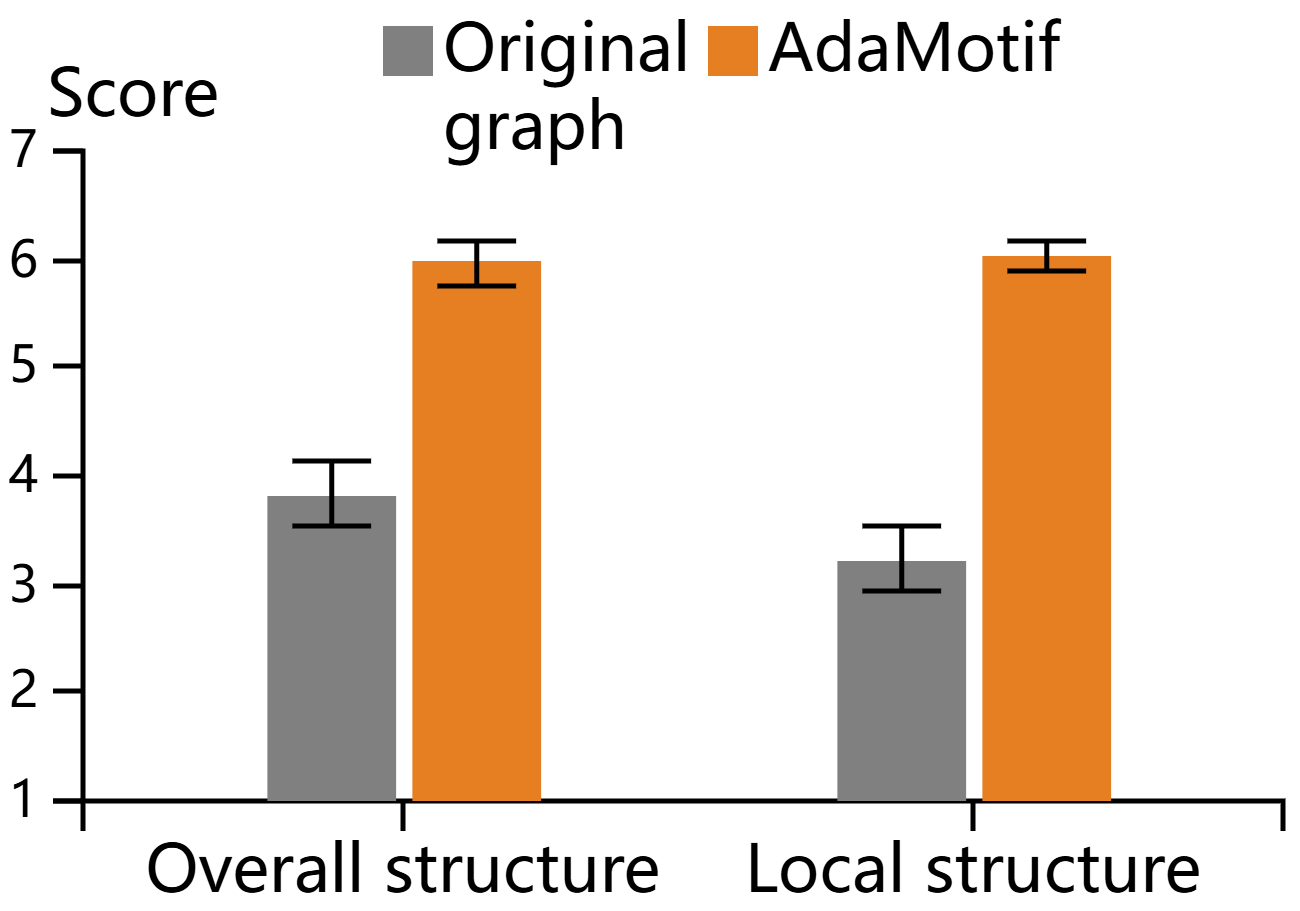}\\
  \caption{Means and standard errors of the Original graph and~\tool{} on overall structure and local structure on a 7-point Likert scale($*, p < 0.05$).
}
  \label{fig:score}
\end{figure}

\section{User Study}
We performed a formal user study to further investigate the effectiveness of the ~\tool{}. The user study compared the \tool{} method with commonly used force-directed node-link diagrams.

\subsection{Hypothesis}

To guide the user study, we formulated four specific hypotheses:

\textbf{H1}. ~\tool{} offers a more robust advantage in finding and analyzing communities. 
Our method reveals that communities with analogous motifs are similar, allowing for clear identification of structurally alike communities within the graph.

\textbf{H2}. ~\tool{} aids in pinpointing distinctions among comparable communities. 
While these communities share motifs, their internal subgraphs reveal unique characteristics.

\textbf{H3}. ~\tool{} facilitates a more intuitive visual estimation of the graph's size.
Our method enables an initial approximation of the size of similar subgraphs across different communities, which in turn aids in estimating the overall graph size.

\textbf{H4}. ~\tool{} performs better in understanding both the overall structure (\textbf{H4a}) and local structure (\textbf{H4b}) of the graph. Our method displays multiple communities and showcases both their commonalities and differences, making it easier to be aware of structural information.

\subsection{Study Design}
Our study was structured around a within-subjects design, incorporating both objective and subjective questions.
For the \textbf{\textit{objective}} questions, we employed a Latin-Square sequence to mitigate order effects.
Regarding the \textbf{\textit{subjective}} questions, we conducted semi-structured interviews with identical questions for all the participants. This uniform set of questions was applied across all datasets and methods to maintain consistency.


\textbf{Tasks}. We designed the research tasks and methodology by employing common graph visualization tasks and user rating evaluations. We referred to relevant research methods~\cite{dunne2013motif, abdelaal2022comparative} and validated our hypotheses. Our tasks were multiple-choice or fill-in-the-blank questions, and the accuracy was determined by whether the answers were correct. The specific tasks are as follows:
\begin{enumerate}[wide, labelwidth=!, labelindent=0pt, label*=\arabic*.]
    \item \textbf{\textit{Similarity}}. About how many communities similar to the highlighted community exist in the graph? 
    \item \textbf{\textit{Difference}}. Which community has the largest number of nodes among the four given similar communities in the graph? These four communities are highlighted in the graph with different colors.
    \item \textit{\textbf{Graph Size}}. About how many nodes are in the graph? (Each option is an interval range. The interval ranges are determined based on the scale of the graph.)
    \item \textbf{\textit{Understandability}}. There are four evaluation tasks: assessing the overall comprehensibility of the original graph, the comprehensibility of its communities; the overall comprehensibility of the ~\tool{} graph, and the comprehensibility of its communities. All evaluations are conducted after completing tasks for all tested datasets.
\end{enumerate}

\textbf{Dataset}. We used six datasets which represent three tiers of graph sizes. \textbf{\textit{Small graphs}}: 1) Aves16~\cite{nr}, which has 64 nodes and 177 edges; 2) The character network from \textit{Les Misérables}~\cite{knuth1993stanford}, which has 77 nodes and 254 edges. \textbf{\textit{Medium-sized graphs}}: 3) Bio-diseasome~\cite{nr}, which has 516 nodes and 1188 edges; 4) Cpan~\cite{heymann2009cpan}, which has 839 nodes and 2,127 edges. \textbf{\textit{Large graphs}}: 5) AS-733~\cite{leskovec2005graphs}, which has 6,474 nodes and 13,895 edges; 6) LastFM Asia Social Network~\cite{rozemberczki2020characteristic}, which has 7,624 nodes and 27,806 edges. 

\textbf{Participants}. We recruited 30 participants (19 males and 11 females) for our user study. Among them, 19 participants have a background in computer-related disciplines, whereas the remaining 11 are from various backgrounds in non-computer fields. Additionally, 16 participants have prior knowledge of graphs, in contrast to 14 who are not.
We randomly assigned participants into two groups. Subsequent adjustments were made to balance the groups, ensuring equitable representation across gender, field of study, and familiarity with graphs. This was done to meet the requirements of the within-subjects design. 

\textbf{Procedure}. Each participant took approximately 30 minutes to complete the tasks. With each dataset divided into original and ~\tool{} graphs, and with the same tasks for each, there are a total of 36 questions across the six datasets. After completing the dataset tasks, there are four evaluation tasks, making it a total of 40 tasks per participant. We set a timer for the first 36 dataset tasks to assess the speed of participants' responses. Two additional, simple tasks involving counting only a limited number of nodes ($\leqslant 8$) and communities ($\leqslant 4$) served as an attention check for participant attentiveness. A reward mechanism was implemented to incentivize focused participation,~\revise{based on the local minimum hourly wage standards (i.e., 30 RMB per hour in Shenzhen, China).} 

\subsection{Result}

We conducted statistical analysis for each graph size separately, taking into account both the original and simplified graphs.
We tracked the accuracy and the duration required by users to complete the tasks. At the smallest graph dimensions, our method excelled in the \textit{similarity} task, while its performance in the remaining tasks aligned closely with that of the original graphs.
However, as the scale increased, the advantages of our method became more pronounced. Details are in Table.~\ref{tab:result}. Additionally, we conducted a statistical analysis across all data sizes and tasks to reflect their performance, which we represented using error bars. Details are in Fig.~\ref{fig:task}. Overall, the ~\tool{} outperformed the original graph across all tasks. Consistently, our method obtained better results in \textit{understandability} tasks. Details are in Fig.~\ref{fig:score}. 

We conducted an analysis of the obtained statistical results. For each task, we performed measures analysis of means and standard errors on the collected data. For measures with statistically significant differences, we employed the following test protocol: Given the use of non-standardized scales, it was imperative to first ascertain the normality of the data distribution.
We performed the Shapiro-Wilk test for each task's data to verify 
normal distribution adherence.
In cases of normal distribution, an independent-samples t-test was utilized for metrics with significant differences. 
Conversely, for data not conforming to normal distribution, the Wilcoxon signed-rank test was applied.
The specifics of normal distribution test are detailed in Table.~\ref{tab:Normal1} and~\ref{tab:Normal2}. 

\textbf{Similarity}. 
In the study, participants demonstrated a higher accuracy for using \tool{} ($Mean_{acc} = 0.845$, $SD_{acc} = 0.048$) to identify structurally similar communities within graphs compared to the original graph ($Mean_{acc} = 0.344$, $SD_{acc} = 0.092$). The accuracy score of \tool{} was statistically significantly higher by 0.500, 95\% CI [$0.335~to~0.666$] than the original graph, $t(4)_{acc} = 8.381$, $p_{acc} < 0.001$, via the t-test.
In terms of task completion time, \tool{} ($Mean_{time} = 24.417$, $SD_{time} = 30.709$) also outperformed the original graph ($Mean_{time} = 36.060$, $SD_{time} = 30.177$). Based on the Wilcoxon signed-rank test, there was a significant difference between our method and the original graph ($Z_{time} = -5.257, p_{time} < 0.001$).
Both in terms of accuracy and time, ~\tool{} showed significantly superior performance (Table.~\ref{tab:result}, Fig.~\ref{fig:task}, \textbf{H1 accepted}). 

\textbf{Difference}. Participants performed well utilizing both \tool{} and the original graph in telling different communities.
The accuracy results for \tool ($Mean_{acc} = 0.750$, $SD_{acc} = 0.109$) showed a bit higher than the original graph ($Mean_{acc} = 0.717$, $SD_{acc} = 0.137$). Statistical analysis via the t-test revealed no significance with the mean difference of 0.033, 95\% CI [$-0.247~to~0.314$], $t(4)_{acc} = 0.330$, $p_{acc} = 0.379$. 
In terms of time efficiency, \tool{} ($Mean_{time} = 20.671$, $SD_{time} = 21.311$) recorded slightly faster than the original graph ($Mean_{time} = 26.172$, $SD_{time} = 26.851$). Based on the Wilcoxon signed-rank test, there was a significant difference between our method and the original graph ($Z_{time} = -3.283, p_{time} = 0.001$).
Our method achieves slightly better accuracy than the original graph in distinguishing similar communities (Table.~\ref{tab:result}, Fig.~\ref{fig:task}), while also being more efficient in handling these tasks.
(\textbf{H2 partially accepted}). 

\textbf{Graph Size}. 
Visual estimation of node counts in graphs becomes increasingly challenging with larger datasets.
The use of \tool{} ($Mean_{acc} = 0.511$, $SD_{acc} = 0.220$) showed to enhance the accuracy of such estimations when compared with the original graph ($Mean_{acc} = 0.361$, $SD_{acc} = 0.168$).
There was no statistically significant mean difference of 0.150, 95\% CI [$-0.294~to~0.594$], $t(4)_{acc} = 0.940$, $p_{acc} = 0.200$.
In terms of time period assumption, there is no improvement of \tool ($Mean_{time} = 37.239$, $SD_{time} = 40.734$) over the original graph ($Mean_{time} = 33.440$, $SD_{time} = 29.904$). Based on the Wilcoxon signed-rank test, there was no significant difference between our method and the original graph ($Z_{time} = -0.996, p_{time} = 0.319$).
Our method still faces challenges in performance with very large datasets (Table.~\ref{tab:result}, Fig.~\ref{fig:task}, \textbf{H3 rejected}).


\textbf{Understandability}. Our method ($Mean = 5.965$, $SD = 1.048$) significantly outperforms the original graph ($Mean = 3.835$, $SD = 1.551$) in terms of overall structural comprehensibility (Fig.~\ref{fig:score}). Based on the Wilcoxon signed-rank test, there was a significant difference between our method and the original graph ($Z = 35.500, p < 0.05$, \textbf{H4a accepted}). Similarly, in terms of local structural comprehensibility, our method ($Mean = 6.035$, $SD = 0.795$) also outperforms the original graph ($Mean = 3.235$, $SD = 1.542$) (Fig.~\ref{fig:score}).
The Wilcoxon signed-rank test suggests that there was a significant difference between our method and the original graph ($Z = 5.000, p < 0.05$, \textbf{H4b accepted}).

In summary, our method outperforms the original graph across multiple evaluation tasks. Particularly, in the case of small-scale data, our method excels in identifying similar community structures. As the data scale increases, the advantages of our method become even more pronounced, demonstrating its effectiveness in graph simplification. Both user ratings and verbal feedback highlight the strong performance of our method. The adaptive motifs enable users to gain a deeper understanding of each community and analyze the required information more efficiently. Therefore, our method effectively simplifies the graph.

\section{Discussion and Future Work}


The experiments demonstrate that \tool{} possesses distinct benefits. 
Our simplified results are aesthetically appealing and significantly reduce visual clutter while accentuating different patterns of graph communities. 
Our method was initially inspired by the motif simplification approach~\cite{dunne2013motif} which can discover and represent only three specific types of structures: fan, clique, and connector. On the contrary, our method is applicable to general graph structures. Furthermore, our method can automatically discover any type of structure in the graph and represent it with adaptive motifs. Our method currently only works on undirected graphs. In the future, we plan to study the simplification of other types of graphs (e.g., directed graphs and dynamic graphs).

While the communities revealed by our motifs are highly dependent on the selected community detection algorithm~\cite{blondel2008fast}, it is not limited to that algorithm because each step of our framework is relatively independent, and replacing that algorithm with other community detection method will not affect the whole framework. Therefore, theoretically, users can choose any algorithm based on their requirements to discover the communities and represent them as motifs of different shapes.

The community similarities and differences reflected in our motifs are based on the graph clustering and graph alignment algorithms utilized. Different graph clustering algorithms may generate different results, and the graph similarities and differences identified by graph alignment algorithms may not be accurate. Although we know that the maximum common subgraph method will give more accurate results, it is an NP-hard problem, which severely limits the efficiency of our method on large graphs. Therefore, we choose to use the graph alignment method with lower time complexity and higher inaccuracy. In the future, we plan to explore whether there is any heuristic maximum common subgraph method to improve the accuracy of our method. 

Our method is more effective for graphs with clear community structures. When the graph
does not own clear communities, the advantage of our approach in visualizing large graphs can be weakened.
Furthermore, our method is effective for large graphs, it is important to note that our implemented algorithms are still limited by the computer memory and screen size.
For extreme large graphs, our approach can still suffer from scalability issues in terms of computation time and visual clutter. 
In the future, we will improve the scalability of our method to handle larger graphs. One possibility is to use GPU-accelerated algorithms to handle super-large graphs and design methods for multi-layered displays to address the issue of small screens.

In the field of graph simplification, information loss is an inevitable challenge. Our method loses the original edge information between nodes across communities. Therefore, we plan to design an appropriate encoding scheme for the simplified edges in our results in the future. 

For the layout of different community motifs, we use the basic force-directed layout algorithm~\cite{d3-component} in our current implementation. It will be interesting to incorporate other algorithms~\cite{xue2022taurus, archambault2007topolayout} into our approach to further enhance our motif layout effect, which is left as future work.



\section{Conclusion}

\revise{We have proposed a novel framework for simplifying large graphs using adaptive motif design, which combines the advantages of graph visualization and graph mining techniques. Communities in the graph are first detected, represented as subgraphs, and then clustered. A representative subgraph is computed for each category. Based on graph alignment results, a similarity-aware representative subgraph layout algorithm, and a difference-aware individual subgraph layout algorithm are used to decide the subgraph layouts.} Finally, adaptive motifs are generated, and these motifs replace the original subgraphs. Thus, our method significantly simplifies the original graph, making it easier for users to analyze and explore the community structures.


 \acknowledgments{%
 The authors would like to thank all the participants for their
participation in the user studies and interviews, Dr. Shu Yang for
the constructive discussions, and all the anonymous reviewers for
their valuable comments and suggestions. This work is partially supported by the Shenzhen Science and Technology Major Project (KJZD20230923114605011), the Scientific Development Funds of Shenzhen University (No. 000001032518), the National Natural Science Foundation of China (No. 62273241), the Natural Science Foundation of Guangdong Province, China (No. 2024A1515011946), and NTU Start Up Grant.
 }

\section*{Supplemental Materials}
\label{sec:supplemental_materials}

All supplemental materials are available on OSF at \url{https://osf.io/pb8t3/}.
In particular, these include: (1) datasets used to generate the figures in the paper and the user study examples, (2) figure images in multiple formats, and (3) all materials related to the user study. 
Our code is shared at \url{https://github.com/lpfeng11/AdaMotif}.

\section*{Figure Credits}
\label{sec:figure_credits}

Fig.~\ref{fig:workflow}(a) is generated by using the force-directed graph layout component of D3~\cite{d3-component}.


\bibliographystyle{abbrv-doi-hyperref}

\bibliography{template}

\begin{thebibliography}{10}

\bibitem{polybuffer}
https://ww2.mathworks.cn/help/matlab/ref/polyshape.polybuffer.html.

\bibitem{abdelaal2022comparative}
M.~Abdelaal, N.~D. Schiele, K.~Angerbauer, K.~Kurzhals, M.~Sedlmair, and D.~Weiskopf.
\newblock Comparative evaluation of bipartite, node-link, and matrix-based network representations.
\newblock {\em IEEE Transactions on Visualization and Computer Graphics}, 29(1):896--906, 2022.
\newblock doi: 10.1109/TVCG.2022.3209427.

\bibitem{alon2007network}
U.~Alon.
\newblock Network motifs: theory and experimental approaches.
\newblock {\em Nature Reviews Genetics}, 8(6):450--461, 2007.
\newblock doi: 10.1038/nrg2102.

\bibitem{archambault2007topolayout}
D.~Archambault, T.~Munzner, and D.~Auber.
\newblock Topolayout: Multilevel graph layout by topological features.
\newblock {\em IEEE transactions on visualization and computer graphics}, 13(2):305--317, 2007.
\newblock doi: 10.1109/TVCG.2007.46.

\bibitem{blondel2008fast}
V.~D. Blondel, J.-L. Guillaume, R.~Lambiotte, and E.~Lefebvre.
\newblock Fast unfolding of communities in large networks.
\newblock {\em Journal of statistical mechanics: theory and experiment}, 2008(10):P10008, 2008.
\newblock doi: 10.1088/1742-5468/2008/10/P10008.

\bibitem{brehmer2021generative}
M.~Brehmer, R.~Kosara, and C.~Hull.
\newblock Generative design inspiration for glyphs with diatoms.
\newblock {\em IEEE Transactions on Visualization and Computer Graphics}, 28(1):389--399, 2021.
\newblock doi: 10.1109/TVCG.2021.3114792.

\bibitem{cabrera2021influence}
B.~Cabrera, B.~Ross, D.~R{\"o}chert, F.~Br{\"u}nker, and S.~Stieglitz.
\newblock The influence of community structure on opinion expression: an agent-based model.
\newblock {\em Journal of Business Economics}, 91:1331--1355, 2021.
\newblock doi: 10.1007/s11573-021-01064-7.

\bibitem{chen2020cone}
X.~Chen, M.~Heimann, F.~Vahedian, and D.~Koutra.
\newblock Cone-align: Consistent network alignment with proximity-preserving node embedding.
\newblock In {\em Proceedings of the 29th ACM International Conference on Information \& Knowledge Management}, pp. 1985--1988, 2020.
\newblock doi: 10.1145/3340531.3412136.

\bibitem{collins2009bubble}
C.~Collins, G.~Penn, and S.~Carpendale.
\newblock Bubble sets: Revealing set relations with isocontours over existing visualizations.
\newblock {\em IEEE transactions on visualization and computer graphics}, 15(6):1009--1016, 2009.
\newblock doi: 10.1109/TVCG.2009.122.

\bibitem{coupette2021graph}
C.~Coupette and J.~Vreeken.
\newblock Graph similarity description: How are these graphs similar?
\newblock In {\em Proceedings of the 27th ACM SIGKDD Conference on Knowledge Discovery \& Data Mining}, pp. 185--195, 2021.
\newblock doi: 10.1145/3447548.3467257.

\bibitem{cui2022survey}
Y.~Cui, X.~Li, J.~Li, H.~Wang, and X.~Chen.
\newblock A survey of sampling method for social media embeddedness relationship.
\newblock {\em ACM Computing Surveys}, 55(4):1--39, 2022.
\newblock doi: 10.1145/3524105.

\bibitem{d3-shape}
D3.
\newblock https://github.com/d3/d3-shape\#curvebundle.

\bibitem{d3-component}
D3.
\newblock https://observablehq.com/@d3/force-directed-graph-component.

\bibitem{de2011generalized}
P.~De~Meo, E.~Ferrara, G.~Fiumara, and A.~Provetti.
\newblock Generalized louvain method for community detection in large networks.
\newblock In {\em 2011 11th international conference on intelligent systems design and applications}, pp. 88--93. IEEE, 2011.
\newblock doi: 10.1109/ISDA.2011.6121636.

\bibitem{doncheva2012topological}
N.~T. Doncheva, Y.~Assenov, F.~S. Domingues, and M.~Albrecht.
\newblock Topological analysis and interactive visualization of biological networks and protein structures.
\newblock {\em Nature protocols}, 7(4):670--685, 2012.
\newblock doi: 10.1038/nprot.2012.004.

\bibitem{dunne2013motif}
C.~Dunne and B.~Shneiderman.
\newblock Motif simplification: improving network visualization readability with fan, connector, and clique glyphs.
\newblock In {\em Proceedings of the SIGCHI Conference on Human Factors in Computing Systems}, pp. 3247--3256, 2013.
\newblock doi: 10.1145/2470654.2466444.

\bibitem{edelsbrunner1983shape}
H.~Edelsbrunner, D.~Kirkpatrick, and R.~Seidel.
\newblock On the shape of a set of points in the plane.
\newblock {\em IEEE Transactions on information theory}, 29(4):551--559, 1983.
\newblock doi: 10.1109/TIT.1983.1056714.

\bibitem{frey2007clustering}
B.~J. Frey and D.~Dueck.
\newblock Clustering by passing messages between data points.
\newblock {\em science}, 315(5814):972--976, 2007.
\newblock doi: 10.1126/science.1136800.

\bibitem{fuchs2016systematic}
J.~Fuchs, P.~Isenberg, A.~Bezerianos, and D.~Keim.
\newblock A systematic review of experimental studies on data glyphs.
\newblock {\em IEEE transactions on visualization and computer graphics}, 23(7):1863--1879, 2016.
\newblock doi: 10.1109/TVCG.2016.2549018.

\bibitem{gao2010survey}
X.~Gao, B.~Xiao, D.~Tao, and X.~Li.
\newblock A survey of graph edit distance.
\newblock {\em Pattern Analysis and applications}, 13:113--129, 2010.
\newblock doi: 10.1007/s10044-008-0141-y.

\bibitem{grover2016node2vec}
A.~Grover and J.~Leskovec.
\newblock node2vec: Scalable feature learning for networks.
\newblock In {\em Proceedings of the 22nd ACM SIGKDD international conference on Knowledge discovery and data mining}, pp. 855--864, 2016.
\newblock doi.org/10.1145/2939672.2939754.

\bibitem{guo2020survey}
Q.~Guo, F.~Zhuang, C.~Qin, H.~Zhu, X.~Xie, H.~Xiong, and Q.~He.
\newblock A survey on knowledge graph-based recommender systems.
\newblock {\em IEEE Transactions on Knowledge and Data Engineering}, 34(8):3549--3568, 2020.
\newblock doi: 10.1109/TKDE.2020.3028705.

\bibitem{gupta2020developing}
A.~Gupta and R.~K. Singh.
\newblock Developing a framework for evaluating sustainability index for logistics service providers: graph theory matrix approach.
\newblock {\em International Journal of Productivity and Performance Management}, 69(8):1627--1646, 2020.
\newblock doi: 10.1108/IJPPM-12-2019-0593.

\bibitem{hartigan1979algorithm}
J.~A. Hartigan and M.~A. Wong.
\newblock Algorithm as 136: A k-means clustering algorithm.
\newblock {\em Journal of the royal statistical society. series c (applied statistics)}, 28(1):100--108, 1979.
\newblock doi: 10.2307/2346830.

\bibitem{hartmanis1982computers}
J.~Hartmanis.
\newblock Computers and intractability: a guide to the theory of np-completeness (michael r. garey and david s. johnson).
\newblock {\em Siam Review}, 24(1):90, 1982.
\newblock doi: 10.1137/1024022.

\bibitem{heer2005vizster}
J.~Heer and D.~Boyd.
\newblock Vizster: Visualizing online social networks.
\newblock In {\em IEEE Symposium on Information Visualization, 2005. INFOVIS 2005.}, pp. 32--39. IEEE, 2005.
\newblock doi: 10.1109/INFVIS.2005.1532126.

\bibitem{heimann2021refining}
M.~Heimann, X.~Chen, F.~Vahedian, and D.~Koutra.
\newblock Refining network alignment to improve matched neighborhood consistency.
\newblock In {\em Proceedings of the 2021 SIAM International Conference on Data Mining (SDM)}, pp. 172--180. SIAM, 2021.
\newblock doi: 10.1137/1.9781611976700.20.

\bibitem{hermanns2023grasp}
J.~Hermanns, K.~Skitsas, A.~Tsitsulin, M.~Munkhoeva, A.~Kyster, S.~Nielsen, A.~M. Bronstein, D.~Mottin, and P.~Karras.
\newblock Grasp: Scalable graph alignment by spectral corresponding functions.
\newblock {\em ACM Transactions on Knowledge Discovery from Data}, 17(4):1--26, 2023.
\newblock doi: 10.1145/3561058.

\bibitem{heymann2009cpan}
S.~Heymann.
\newblock Cpan-explorer, an interactive exploration of the perl ecosystem. https://gephi.wordpress.com/2009/06/25/cpan-explorer-an-interactive-exploration-of-the-perl-ecosystem/.
\newblock {\em Gephi Blog}, 2009.

\bibitem{hoeber2018information}
O.~Hoeber.
\newblock Information visualization for interactive information retrieval.
\newblock In {\em Proceedings of the 2018 Conference on Human Information Interaction \& Retrieval}, pp. 371--374, 2018.
\newblock doi: 10.1145/3176349.3176898.

\bibitem{holten2009force}
D.~Holten and J.~J. Van~Wijk.
\newblock Force-directed edge bundling for graph visualization.
\newblock In {\em Computer graphics forum}, vol.~28, pp. 983--990. Wiley Online Library, 2009.
\newblock doi: 10.1111/j.1467-8659.2009.01450.x.

\bibitem{jiao2022hierarchical}
B.~Jiao, X.~Lu, J.~Xia, B.~B. Gupta, L.~Bao, and Q.~Zhou.
\newblock Hierarchical sampling for the visualization of large scale-free graphs.
\newblock {\em IEEE Transactions on Visualization and Computer Graphics}, 2022.
\newblock doi: 10.1109/TVCG.2022.3201567.

\bibitem{kann1992approximability}
V.~Kann.
\newblock On the approximability of the maximum common subgraph problem.
\newblock In {\em STACS 92: 9th Annual Symposium on Theoretical Aspects of Computer Science Cachan, France, February 13--15, 1992 Proceedings 9}, pp. 375--388. Springer, 1992.
\newblock doi: 10.1007/3-540-55210-3\_198.

\bibitem{knuth1993stanford}
D.~E. Knuth.
\newblock {\em The Stanford GraphBase: a platform for combinatorial computing}, vol.~1.
\newblock AcM Press New York, 1993.

\bibitem{koutra2015summarizing}
D.~Koutra, U.~Kang, J.~Vreeken, and C.~Faloutsos.
\newblock Summarizing and understanding large graphs.
\newblock {\em Statistical Analysis and Data Mining: The ASA Data Science Journal}, 8(3):183--202, 2015.
\newblock doi: 10.1002/sam.11267.

\bibitem{lee2020ssumm}
K.~Lee, H.~Jo, J.~Ko, S.~Lim, and K.~Shin.
\newblock Ssumm: Sparse summarization of massive graphs.
\newblock In {\em Proceedings of the 26th ACM SIGKDD International Conference on Knowledge Discovery \& Data Mining}, pp. 144--154, 2020.
\newblock doi: 10.1145/3394486.3403057.

\bibitem{leskovec2005graphs}
J.~Leskovec, J.~Kleinberg, and C.~Faloutsos.
\newblock Graphs over time: densification laws, shrinking diameters and possible explanations.
\newblock In {\em Proceedings of the eleventh ACM SIGKDD international conference on Knowledge discovery in data mining}, pp. 177--187, 2005.
\newblock doi: 10.1145/1081870.1081893.

\bibitem{liu2018graph}
Y.~Liu, T.~Safavi, A.~Dighe, and D.~Koutra.
\newblock Graph summarization methods and applications: A survey.
\newblock {\em ACM computing surveys (CSUR)}, 51(3):1--34, 2018.
\newblock doi: 10.1145/3186727.

\bibitem{majeed2020social}
S.~Majeed, M.~Uzair, U.~Qamar, and A.~Farooq.
\newblock Social network analysis visualization tools: A comparative review.
\newblock In {\em 2020 IEEE 23rd International Multitopic Conference (INMIC)}, pp. 1--6. IEEE, 2020.
\newblock doi: 10.1109/INMIC50486.2020.9318162.

\bibitem{narayanan2017graph2vec}
A.~Narayanan, M.~Chandramohan, R.~Venkatesan, L.~Chen, Y.~Liu, and S.~Jaiswal.
\newblock graph2vec: Learning distributed representations of graphs.
\newblock {\em arXiv preprint arXiv:1707.05005}, 2017.
\newblock doi: 10.48550/arXiv.1707.05005.

\bibitem{nassar2018low}
H.~Nassar, N.~Veldt, S.~Mohammadi, A.~Grama, and D.~F. Gleich.
\newblock Low rank spectral network alignment.
\newblock In {\em Proceedings of the 2018 World Wide Web Conference}, pp. 619--628, 2018.
\newblock doi: 10.1145/3178876.3186128.

\bibitem{NodeXL}
NodeXL.
\newblock https://nodexl.com/.

\bibitem{oliver2023scalable}
P.~Oliver, E.~Zhang, and Y.~Zhang.
\newblock Scalable hypergraph visualization.
\newblock {\em IEEE Transactions on Visualization and Computer Graphics}, 2023.
\newblock doi: 10.1109/TVCG.2023.3326599.

\bibitem{purohit2014fast}
M.~Purohit, B.~A. Prakash, C.~Kang, Y.~Zhang, and V.~Subrahmanian.
\newblock Fast influence-based coarsening for large networks.
\newblock In {\em Proceedings of the 20th ACM SIGKDD international conference on Knowledge discovery and data mining}, pp. 1296--1305, 2014.
\newblock doi: 10.1145/2623330.2623701.

\bibitem{raynor2022state}
J.~Raynor, T.~Crnovrsanin, S.~Di~Bartolomeo, L.~South, D.~Saffo, and C.~Dunne.
\newblock The state of the art in bgp visualization tools: A mapping of visualization techniques to cyberattack types.
\newblock {\em IEEE Transactions on Visualization and Computer Graphics}, 29(1):1059--1069, 2022.
\newblock doi: 10.1109/TVCG.2022.3209412.

\bibitem{nr}
R.~A. Rossi and N.~K. Ahmed.
\newblock The network data repository with interactive graph analytics and visualization.
\newblock In {\em AAAI}, 2015.
\newblock doi: 10.1609/aaai.v29i1.9277.

\bibitem{rozemberczki2020characteristic}
B.~Rozemberczki and R.~Sarkar.
\newblock Characteristic functions on graphs: Birds of a feather, from statistical descriptors to parametric models.
\newblock In {\em Proceedings of the 29th ACM international conference on information \& knowledge management}, pp. 1325--1334, 2020.
\newblock doi: 10.1145/3340531.3411866.

\bibitem{tabassum2018social}
S.~Tabassum, F.~S. Pereira, S.~Fernandes, and J.~Gama.
\newblock Social network analysis: An overview.
\newblock {\em Wiley Interdisciplinary Reviews: Data Mining and Knowledge Discovery}, 8(5):e1256, 2018.
\newblock doi: 10.1002/widm.1256.

\bibitem{trung2020comparative}
H.~T. Trung, N.~T. Toan, T.~Van~Vinh, H.~T. Dat, D.~C. Thang, N.~Q.~V. Hung, and A.~Sattar.
\newblock A comparative study on network alignment techniques.
\newblock {\em Expert Systems with Applications}, 140:112883, 2020.
\newblock doi: 10.1016/j.eswa.2019.112883.

\bibitem{von2007tutorial}
U.~Von~Luxburg.
\newblock A tutorial on spectral clustering.
\newblock {\em Statistics and computing}, 17:395--416, 2007.
\newblock doi: 10.1007/s11222-007-9033-z.

\bibitem{xue2022taurus}
M.~Xue, Z.~Wang, F.~Zhong, Y.~Wang, M.~Xu, O.~Deussen, and Y.~Wang.
\newblock Taurus: towards a unified force representation and universal solver for graph layout.
\newblock {\em IEEE Transactions on Visualization and Computer Graphics}, 29(1):886--895, 2022.
\newblock doi: 10.1109/TVCG.2022.3209371.

\bibitem{ying2022metaglyph}
L.~Ying, X.~Shu, D.~Deng, Y.~Yang, T.~Tang, L.~Yu, and Y.~Wu.
\newblock Metaglyph: Automatic generation of metaphoric glyph-based visualization.
\newblock {\em IEEE Transactions on Visualization and Computer Graphics}, 29(1):331--341, 2022.
\newblock doi: 10.1109/TVCG.2022.3209447.

\bibitem{zeng2018distributed}
J.~Zeng and H.~Yu.
\newblock A distributed infomap algorithm for scalable and high-quality community detection.
\newblock In {\em Proceedings of the 47th International Conference on Parallel Processing}, pp. 1--11, 2018.
\newblock doi: 10.1145/3225058.3225137.

\bibitem{zhao2020preserving}
Y.~Zhao, H.~Jiang, Y.~Qin, H.~Xie, Y.~Wu, S.~Liu, Z.~Zhou, J.~Xia, F.~Zhou, et~al.
\newblock Preserving minority structures in graph sampling.
\newblock {\em IEEE Transactions on Visualization and Computer Graphics}, 27(2):1698--1708, 2020.
\newblock doi: 10.1109/TVCG.2020.3030428.

\bibitem{zhou2020interactive}
S.~Zhou, X.~Dai, H.~Chen, W.~Zhang, K.~Ren, R.~Tang, X.~He, and Y.~Yu.
\newblock Interactive recommender system via knowledge graph-enhanced reinforcement learning.
\newblock In {\em Proceedings of the 43rd international ACM SIGIR conference on research and development in information retrieval}, pp. 179--188, 2020.
\newblock doi: 10.1145/3397271.3401174.

\bibitem{zhou2022topological}
Y.~Zhou, A.~Rathore, E.~Purvine, and B.~Wang.
\newblock Topological simplifications of hypergraphs.
\newblock {\em IEEE Transactions on Visualization and Computer Graphics}, 2022.
\newblock doi: 10.1109/TVCG.2022.3153895.

\bibitem{zhou2020context}
Z.~Zhou, C.~Shi, X.~Shen, L.~Cai, H.~Wang, Y.~Liu, Y.~Zhao, and W.~Chen.
\newblock Context-aware sampling of large networks via graph representation learning.
\newblock {\em IEEE Transactions on Visualization and Computer Graphics}, 27(2):1709--1719, 2020.
\newblock doi: 10.1109/TVCG.2020.3030440.

\end{thebibliography}

\appendix 

\end{document}